\documentstyle[12pt]{article}
\begin{document}  
\renewcommand{\thefootnote}{\fnsymbol{footnote}}
\renewcommand{\theequation}{\arabic{section}.\arabic{equation}}  
\font \tensbold                = cmssbx10  
\font \sevensbold              = cmssbx10 at 7pt  
\font \fivesbold               = cmssbx10 at 5pt  
\newfam\sboldfam  
\textfont\sboldfam=\tensbold \scriptfont\sboldfam=\sevensbold  
\scriptscriptfont\sboldfam=\fivesbold  
\def\sbold{\fam\sboldfam\tensbold}
\parskip = 0 pt  
\parindent = 10 pt  
\abovedisplayskip=13pt plus 3pt minus 9pt
\belowdisplayskip=13pt plus 3pt minus 9pt
\centerline{\bf SUBDYNAMICS THROUGH TIME SCALES AND SCATTERING MAPS}\par  
\centerline{\bf IN QUANTUM FIELD THEORY}\par
\vskip 5 pt  
\centerline{Ludovico~Lanz$^{(1)}$, Olaf~Melsheimer$^{(2)}$ and  
Bassano~Vacchini$^{(1)}$}\par
{\it  
\centerline{$^{(1)}$Dipartimento di Fisica dell'Universit\`a di  
Milano, INFN, Sezione di Milano,}  
\centerline{Via Celoria 16,  
I-20133, Milano, Italy}  
\centerline{$^{(2)}$Fachbereich Physik, Philipps-Universit\"at, Renthof 7,  
D-3550  Marburg, Germany}  
}  
\vskip 10 pt  
\centerline{\sc Abstract}\par  
{ \baselineskip=11pt  \it 
It is argued that the dynamics of an isolated system, due to the concrete 
procedure by which it is separated from the environment, has a non-Hamiltonian 
contribution. By a  unified quantum field theoretical treatment  of typical 
subdynamics, e.g., hydrodynamics, kinetic theory, master equation
for a particle
interacting with matter, we look for the  structure of this more general 
dynamics.                                     
}\par  
\vskip 10 pt  
\noindent  
\par  
\section{The concept of physical system}  
\par         
Quantum mechanics (QM) has non-separability as its most  
striking feature, i.e., one  
cannot attribute ``properties'' to parts of a system and therefore typical  
problems like the measurement process and EPR situations  arise.  
This feature is so deeply rooted in the mathematical structure of QM that we  
believe one should not try to make it less stringent, e.g., by attempts like
``spontaneous reduction''~\cite{GRW}.  
 We prefer instead to weaken the very concept of physical system: usually the  
``isolation'' of a physical system is taken for granted, while in our opinion  
the way in which isolation is achieved belongs to the  very definition of   
the system.  
Any attempt inside QM to obtain the subdynamics for a subsystem  
enforces the introduction of a suitable time scale in order to break the  
correlations with  the environment; in a completely sharp  
description of the dynamics of a subsystem the physics of the whole universe  
would enter.  
The preparation procedure leading to a system, isolated during a time interval  
$[t_0,t_1]$ and confined in a spatial region $\omega$, covers a time interval  
$[T,t_0]$, that will be called ``preparation time''.  
Due to the confinement the basic space-time symmetries are broken and by  
suitable boundary conditions ``peculiar'' properties of the system are  
introduced. This obviously reduces the universal character of the dynamical  
description, however an important universal behaviour still remains due to  
symmetry and locality (or short range character) of effective  
interactions, whose relevance becomes particularly evident in the  
quantum field theoretical approach. We thus regard a physical  
system as a part of the  
world under control by a suitable preparation, whose  
local behaviour is explained  
in terms of locally interacting quantum fields.  
The choice of these fields depends on the level of description of the  
system. A large part of physics can be explained in terms of quantum fields  
related to molecules with a typical time scale of the order  
$\approx 10^{-13}$ s;  
 a much more refined description arises if the basic fields are related to  
nuclei and electrons, then the basic theory would be QED and a much smaller  
time scale $\approx 10^{-23}$ s could be considered: however such
role of QED as  
basic theory of macrosystems is far from being exploited.  
\par 
In a  
sense our viewpoint  
can appear as opposite to the most widely spread one, which we  
synthetise as  
follows: particles are the primary systems, related to
non-confined quantum fields and to basic symmetries, all other systems are  
structure of particles; one then  
tries to obtain a typical macroscopic behaviour in  
some suitable thermodynamic limit. According to us, on the  
contrary,  
macroscopic  
systems are to be taken as  
the primary systems, even if in their definition time scales and  
spatial confinement must be carefully taken into account; the theoretical  
framework for their description is quantum field theory, locality and  
quantisation taking the place of the atomistic model. In this context
particles  
are a derived concept.  
The description of non-equilibrium systems  
is put in the foreground and  
at least in principle should be performed taking boundary effects  
into account; procedures like ``continuous'' limit should be  
applied only at the end, if one wants to get rid of boundary  
effects.  
This standpoint is closer to thermodynamics and  
electromagnetism, while the former one originates from classical mechanics.
The  
relevance of macroscopic systems for the foundations of QM is the starting  
point of Ludwig's axiomatic approach to QM~\cite{Foundations}.  
The insistence on the distinction between these two attitudes is  
due to the fact that they lead  
in a natural  
way to two different formulations of the dynamics.  
In the first approach one associates a wave function $\psi$ to  
each system  
 ($\psi ({\mbox {\bf x}},t)$ for one particle,  
$\ldots\,$,  
$\psi ({\mbox {\bf x}}_1,{\mbox {\bf x}}_2,\ldots,{\mbox {\bf x}}_N,t)$
for  N particles);  
obviously if one describes  
situations like ``unpolarised'' particles it is appropriate to use a  
statistical operator, in order to take a lack of control of the  
experimental specification into account.  
This aspect becomes increasingly  
important for large N, so that statistical operators are very useful for  
macroscopic systems, nevertheless the basic dynamics is given by  
an evolution operator  
for the wavefunction  $\psi$.  
On the other hand,  
starting with a macroscopic system,  
one is led to assume  
a statistical operator ${\hat \varrho}_t$  as the most appropriate  
mathematical  
representation of the preparation procedure until time $t$.   
The set $\cal K({\cal H})$ of  
statistical operators on the Hilbert space ${\cal H}$  
becomes most important and the space $\cal T ({\cal H}) $ of trace-class  
operators,  
which is generated in a natural way by $\cal K({\cal H})$  [$\cal K({\cal H})$  
is the base of  
the base-norm space $\cal T ({\cal H})$], plays a role similar to  
that of ${\cal H}$  
in the previous formalism. Correspondingly unitary operators on ${\cal H}$  
in the first approach are replaced by affine  
maps of $\cal K({\cal H})$  in   $\cal K({\cal H})$,  
i.e. by positive, trace-preserving maps on $\cal T ({\cal H})$.  
If the  system is isolated in the time interval  
$[t_0,t_1]$ the spontaneous repreparations ${\hat \varrho}_t$,  
$t \in  
[t_0,t_1]$, are related together by  
$\hat \varrho_t = {\cal M}_{t t'} \hat \varrho_{t'}$ ($t\geq t'$),  
where the evolution  system $  
\left \{  
{{\cal M}}_{t'' t'} \,  
t''\geq t'  
\right \}  
$ satisfies the composition rule   
${\cal M}_{t''' t'}={\cal M}_{t''' t''}{\cal M}_{t'' t'}$  
 ($t'\leq t''\leq t'''$).  
We stress the fact that there is no reason to assume that  
${\cal M}_{t'' t'}$ has an inverse. If ${\cal M}^{-1}_{t'' t'}$  
exists then  
${\cal M}_{t'' t'}=  
        {\hat U}_{t''t'}  
        \cdot  
        {\hat U}^{\scriptscriptstyle\dagger}_{t''t'}  
$ (see for example~\cite{Davies}),  
with ${\hat U}_{t''t'}$ unitary or antiunitary operator and one is  
brought back to the  Hilbert space formalism:  
$        \psi_{t''} = {\hat U}_{t''t'}\psi_{t'}$.  
The dynamics in the present  
framework is indeed more general and has irreversibility as  
typical phenomenon.  
\par 
To determine the maps ${\cal M}_{t'' t'}$ a choice of relevant  
observables is necessary: when a time scale is  
introduced, only those observables should be considered, whose  
expectation values do not appreciably vary in a time interval  
of the order of the time scale. It is thus necessary to  
work in the Heisenberg picture, i.e. with the adjoint map ${\cal  
M}^{'}_{t t_{0}}$, and consider expressions of the form ${\cal  
M}^{'}_{t t_{0}}{\hat A}$, ${\hat A}$ being a relevant observable. For the  
same system different descriptions can be given by different  
choices of relevant observables and corresponding  
time scales: e.g., hydrodynamic or kinetic description of a
continuum.  
Skipping questions of mathematical rigour we  
can assume the differential equation  
        \begin{equation}  
        \label{0}  
        {  
        d  
        \over  
         dt  
        }  
        {\cal M}^{'}_{t t_0} = {\cal L}^{'}_t  
        {\cal M}^{'}_{t t_0}  ,  
        \end{equation}  
and represent ${\cal M}^{'}_{t t_0}$ in the form  
       ${\cal M}^{'}_{t t_0} =  
        T  
        \left(  
        e^{\int_{t_0}^t dt' \, {\cal L}^{'}_{t'}}  
        \right)$ in terms of the generator ${\cal L}_t^{'}$.  
It is well known (rigorously for bounded ${\cal L}^{'}_t$) that if  
${\cal M}^{'}_{t'' t'}$ has the additional property of complete  
positivity (CP), ${\cal L}^{'}_t$ has the Lindblad structure~\cite{Lind}:  
        \begin{equation}  
        {\cal L}^{'}_t{\hat B}= +{i \over \hbar}  
        \left(  
        {{{\hat H}}_t {\hat B}-{\hat B} {{\hat H}}_t}  
        \right)  
         -  
        { {1\over\hbar}}  
        \left(  
        {{\hat A}_t {\hat B} + {\hat B} {\hat A}_t}\right)  
        +{1 \over \hbar} \sum_j  
        {{\hat L}}_{tj}^{{\scriptscriptstyle \dagger}}  
        {\hat B}  
        {{\hat L}}_{tj}  
        \label{1}  
        \end{equation}  
        \[  
        {{\hat H}}_t ={{\hat H}}_t^{\scriptscriptstyle \dagger}  
        \qquad  
        {\hat A}_t =  {1\over2}     \sum_j  
        {{\hat L}}_{tj}  {{\hat L}}_{tj}^{{\scriptscriptstyle \dagger}}.  
        \]  
In our framework  
the assumption of CP can appear too restrictive since only a suitable subset  
of observables is relevant and one  
expects  
that a modified concept of CP of ${\cal  
M}^{'}_{t'' t'}$ relatively to these observables should be  
given,  
leading  
to a more general  
structure of ${\cal L}^{'}_t$;  
we shall return to this point in the sequel.  
\par 
The more general description of the dynamics that we are  
considering  
allows the introduction of 
the concept of  
trajectory  in quantum theory.  
In fact in the general formalism of  
continuous measurement approach~\cite{misure cont}  
one has that  an evolution  
system $  
\{ {\cal M}_{t'' t'} \quad t''\geq t' \}  
$  
with ${\cal L}_t$ having the Lindblad structure can be  
decomposed on a space $Y^{t_1}_{t_0}$ of trajectories for  
stochastic variables $\xi(t)$.   
The concept of  
trajectory  
is particularly useful in the field of quantum optics, 
as is shown by the many interesting applications to be found in the literature. 
More precisely one can consider Wiener  
 and jump processes related to the operators ${{\hat L}}_{tj}$  
in the sense that the expectations of the increments for example in the case of
jump processes are given by~\cite{Lanz6}:  
	\[  
        \langle  
         d \xi_j(t)  
         \rangle  
         =  
         dt \, {\mbox{\rm Tr}}  
         \left(  
         (  {{\hat L}}^{\scriptscriptstyle\dagger}_{tj}{{\hat L}}_{tj})  
         \hat \varrho (t)  
         \right)  
         .  
	\]  
One can define   
$\sigma$-algebras  
${\cal B}(Y^{t''}_{t'})$ of subsets  
$  
 \Omega^{t''}_{t'}
$                    of             $Y^{t''}_{t'}$  
and  construct operation valued measures  
$  
{\cal F}^{t''}_{t'}( \Omega^{t''}_{t'}  
)$  
on ${\cal B}(Y^{t''}_{t'})$  
in such a way that  
 $       {\cal M}_{t'' t'}  
        =  
        {\cal F}(Y^{t''}_{t'}).  
  $  
Then for any decomposition  
$Y^{t''}_{t'} =  
        \cup_{\alpha}({\Omega^{t''}_{t'}}_{\alpha})  
$  
with disjoint subsets ${\Omega^{t''}_{t'}}_{\alpha}$ one has  
        \begin{equation}  
        \label{4}  
        {\cal M}_{t'' t'}=  
        \sum_\alpha  
        {\cal F}({\Omega^{t''}_{t'}}_{\alpha})  
                 .  
        \end{equation}  
One can therefore claim that the quantum  
dynamics of the  system  is compatible with the evolution of   
classical stochastic variables;  
typically the probability that the trajectory of these  
variables for $t'\leq t \leq t''$ belongs to a subset  
${\Omega^{t''}_{t'}}_{\alpha}$  
is given by  
        $        p({\Omega^{t''}_{t'}}_{\alpha})
        =  
        {\mbox{\rm Tr}}  
        \left(  
        {\cal F}^{t''}_{t'}( {\Omega^{t''}_{t'}}_{\alpha})
        \hat \varrho (t')  
        \right)  
$.  
The decomposition of  
${\cal M}_{t'' t'}$  
by operation valued stochastic processes ${\cal F}^{t''}_{t'}$ on  
a suitable trajectory space  
$Y^{t''}_{t'}$  
is not unique, i.e. there are many compatible objective  
``classical'' pictures which are consistent with the quantum  
evolution, a feature that can be linked with a
``generalised concept of complementarity''.
The very possibility of recovering some kind of classical insight  
into QM is due to  the  
 non-Hamiltonian evolution; obviously  
(\ref{4}) would be inconsistent with  
$        {\cal M}_{t'' t'}=  
        {\hat U}_{t''t'}  
        \cdot  
        {\hat U}^{\scriptscriptstyle\dagger}_{t''t'}  
$, since for  
$\hat \varrho_{t'}=  
|  
\psi_{t'}  
\rangle  
\langle  
\psi_{t'}  
|  
$ the l.h.s. of (\ref{4}) is a pure state and the r.h.s. is a  
mixture.  
\par 
In the framework we have now presented dynamics is given by  
${\cal L}_t^{'}$~\cite{Lanz1};  
while one expects that the Hamiltonian part is  
fixed by the local interactions, the remaining part is connected  
to the preparation procedure of the isolated system: it cannot be  
strictly derived from a Hamiltonian theory of a larger system  
surrounding it (in fact also for this larger  system an ${\cal  
L}_t^{'}$ should be determined). The problem arises to give  
 reasonable criteria for the construction of ${\cal L}_t^{'}$.  
One can expect that  near to equilibrium only  
 the  
 Hamiltonian part   
$ {i\over\hbar} [\hat H, \cdot] $  
is important, as it is clearly indicated by the  
 great success of equilibrium statistical mechanics; however the   
non-Hamiltonian part  of ${\cal L}_t^{'}$ is relevant for  
 irreversible behaviour and             for a full explanation of  
 approach to equilibrium: in fact energy conservation  
 ${\cal L}_t^{'}{\hat H}=0$  (at least on the relevant time scale) grants for  
the existence of an ``eigenstate'' of  ${\cal L}_t^{'}$  with practically zero  
eigenvalue, while one expects that the other eigenvalues       
of  ${\cal L}_t^{'}$ have a negative real part.  
\par  
\section{A  microsystem interacting with matter}  
\par  
We shall take on later the problem of an explicit construction of  
$\hat \varrho (t)$ for a  macrosystem  
${\cal S}_{\rm M}$.  
We assume now that  
$\hat \varrho (t)$  is given  
(e.g., the  system is at equilibrium) and consider the
problem of describing the new  system  
${\cal S} =  
{\cal S}_{\rm M} +{}$ a  microsystem, ${\cal S}$ still being confined inside  
a region $\omega$.  
The Hamiltonian of ${\cal S}$ is:  
        \[  
        {\hat H}={\hat H}_0 + {\hat H}_{\rm M} + {{\hat V}}  
        \qquad  
	{\hat H}_0 = \sum_f  
        {E_f} {{\hat a}^{\scriptscriptstyle \dagger}_{f}} {{\hat a}_{{f}}}   
	\qquad  
         \left[{{{\hat a}_{{f}}},{{\hat a}^{\scriptscriptstyle  
        \dagger}_{g}}}\right]_{\mp}=\delta_{fg}    ,  
        \]  
where $E_f$ are the eigenvalues of the operator  
$  
-{  
\hbar^2  
\over  
       2m  
}  
\Delta_2  
$       :  
        \[  
        -{  
        \hbar^2  
        \over  
        2m  
        }  
        \Delta_2  
        u_f({\mbox{\bf x}})= E_f  
        u_f({\mbox{\bf x}})  
	\qquad  
        u_f({\mbox{\bf x}})=0 \quad {\mbox{\bf x}}\in  
        \partial\omega  
        .  
        \]  
Let us assume for the  statistical operator of ${\cal S}$ the  
following structure:  
        \begin{equation}  
        {\hat \varrho}(t)=  
        \sum_{{g} {f}}{}  
        {{\hat a}^{\scriptscriptstyle \dagger}_{g}}  
        {{\hat \varrho}_{\rm M}}(t) {{\hat a}_{{f}}}  
        {{ \varrho}}_{gf}(t)    ,  
        \label{5}  
        \end{equation}  
        \begin{equation}  
        \label{6}  
        {{\hat a}_{{f}}}{{\hat \varrho}_{\rm M}}=0    .  
        \end{equation}  
The coefficients  
${{\varrho}}_{gf}$ build a positive, trace one  matrix,  
which can be considered as the  
representative of a statistical operator ${{\hat  
\varrho}}^{(1)}(t)$ in the Hilbert space  
${{\cal H}^{(1)}}$                      spanned by the states  
$u_f$      ($  {{\varrho}}_{gf}=  
        \langle  
        u_g |  
        {{\hat  
        \varrho}}^{(1)}  
        | u_f  
        \rangle   $).  
Equation (\ref{6}) indicates that the system ${\cal S}_{\rm M}$ has  
charge  
$  
{\hat Q}=\sum_f {{\hat a}^{\scriptscriptstyle \dagger}_{f}}  
        {{\hat a}_{{f}}}  
$  
with value zero, i.e. it does not contain the  microsystem.  
Equation (\ref{5}) represents the fact that  
${\cal S}_{\rm M}$ has been  perturbed by the additional particle  
and therefore presents a new  
dynamical behaviour contained in the coefficients  
${{ \varrho}}_{gf}(t)$ that can be picked out studying the time  
evolution of the observables  
$        {\hat A}  
        = \sum_{h,k}  
        {\hat a^{\scriptscriptstyle \dagger}_{h}}  
        {A}_{hk}  
        {\hat a_{k}} $;  
in fact the subdynamics of these observables  
provides the QM of a one-particle system with Hilbert space  
${\cal H}^{(1)}$,  statistical operator ${\hat \varrho}^{(1)}$  
and observables ${\hat{\mbox{\sf  A}}}^{(1)}$, with matrix elements  
$  
A_{hk}=  
\langle  
u_h | {\hat{\mbox{\sf  A}}}^{(1)} u_k  
\rangle_{{\cal H}^{(1)}}  
$,  through the formula                         
        \begin{equation}  
        \label{alfa}  
        {\hbox{\rm Tr}}_{{\cal H}}  
        \left(  
        {{\hat A}{\hat \varrho}}(t)  
        \right)  
        =  
        \sum_{hk}  
        {\hbox{\rm Tr}}_{{\cal H}}  
        \left(  
        e^{{i\over \hbar}{\hat H}t}  
        {\hat a}_h^{\scriptscriptstyle\dagger}  
        {\hat a}_k  
        e^{-{i\over \hbar}{\hat H}t}  
        {\hat \varrho}  
        \right)  
        A_{hk}              
         =  
        {\hbox{\rm Tr}}_{{{\cal H}^{(1)}}}  
        \left(  
        {\hat {{\mbox{\sf A}}}^{(1)} {{\hat \varrho}}^{(1)}}(t)  
        \right)     
	.              
	\end{equation}  
One has to study the expression  
$e^{{i\over \hbar}{\hat H}t}  
        {\hat a}_h^{\scriptscriptstyle\dagger}  
        {\hat a}_k  
        e^{-{i\over \hbar}{\hat H}t}$,   
exploiting the fact that the expectation values  
$  
\langle  
{\hat A}  
\rangle_t  
$  
are ``slowly varying'' if the matrix $A_{hk}$ is  
``quasi-diagonal'' (if $A_{hk}=\delta_{hk}$, ${\hat A}$ is a  
conserved charge). We give here only a sketchy account of the  
main points (for details see~\cite{art1}).  
It proves useful to use  in the Heisenberg picture  
a formalism in ${\cal B}({\cal H})$  
reminiscent of usual scattering theory in ${\cal H}$, by means of  
superoperators, typically:  
        \[  
        {\cal H}^{'}={i \over \hbar} [{{\hat H}},\cdot],  \quad  
        {\cal H}^{'}_0={i \over \hbar} [{{\hat H}}_0 + {\hat H}_{\rm  
        M},\cdot],  
        \quad  
        {\cal V}^{'}={i \over \hbar} [{\hat V},\cdot].  
        \]  
In such a context operators of the form  
$  
{\hat a}^{\scriptscriptstyle \dagger}_{h}  
{\hat a}_{k}  
$  
are ``eigenstates'' of  ${\cal  
H}^{'}_0$ with  
eigenvalues  
$  
{i \over \hbar}  
\left(  
E_h - E_k  
\right)  
$.  
Setting ${\cal U}^{'}(t)=e^{{\cal H}^{'}t}$ one has:  
        \begin{eqnarray}  
        \label{7}  
        {{{\cal U}^{'}(t)}}  
        \left(  
        {{{\hat a}^{\scriptscriptstyle \dagger}_{h}}{{\hat a}_{k}}}  
        \right)  
        &=&  
        \left(  
        {{{\cal U}^{'}(t)}}{{\hat a}^{\scriptscriptstyle \dagger}_{h}}  
        \right)  
               \left(  
        {{{{\cal U}^{'}(t)}}{{\hat a}_{k}}}  
               \right)  
        \\  
        &=&  
        {\int\limits_{-i\infty+\eta}^{+i\infty +  \eta}}{  
        dz_1  
        \over  
            2\pi i  
        }     \,     e^{z_1 t}  
        \left(  
        {  
        {1\over{ z_1 - {\cal H}^{'}}}  
        {{\hat a}^{\scriptscriptstyle \dagger}_{h}}}  
              \right)  
        {\int\limits_{-i\infty+\eta}^{+i\infty +  \eta}}{  
        dz_2  
        \over  
            2\pi i  
        }       \,   e^{z_2 t}  
        \left(  
        {  
        {1\over{ z_2 - {\cal H}^{'}}}  
        {{\hat a}_{k}}  
        }  
        \right)  
        \nonumber  
        \end{eqnarray}  
        \begin{equation}  
        \label{8}  
        {{  
        {1\over{ z - {\cal H}^{'}}}  
        }}  
        =  
        {{  
        {1\over{ z - {\cal H}^{'}_0}}  
        }} +{{  
        {1\over{ z - {\cal H}^{'}_0}}  
        }}  
        {\cal T}(z){{  
        {1\over{ z - {\cal H}^{'}_0}}  
        }}  
         \quad \mbox{\rm where} \quad  
        {\cal T}(z)  
        \equiv  
        {\cal V}^{'} + {\cal V}^{'}{{  
        {1\over{ z - {\cal H}^{'}}}  
        }}{\cal V}^{'}      .  
        \end{equation}  
${\cal T}(z)$ is reminiscent of the usual T-matrix of scattering  
theory and plays a central role in this treatment: it will be  
called ``scattering map''. The operator ${\cal T}(z)$ has poles on  
the imaginary axis for $z={i\over\hbar}(e_\alpha-e_\beta)$,  
$e_\alpha$ being the eigenvalues of ${\hat H}$. In the  
calculation of expression (\ref{7}) we shall assume that the  
function ${\cal T}(z)$ for $Re z \approx \varepsilon$ with $  
\varepsilon \gg \delta$ ($\delta$ typical spacing between the  
poles) is smooth enough, so that the only relevant contribution  
from the singularities of $({ z - {\cal H}^{'}})^{-1}$ stems from  
the singularities of $({ z - {\cal H}^{'}_0})^{-1}$; this smoothness  
property is linked to the fact that the set of poles of $({  
z - {\cal H}^{'}})^{-1}$ goes over to a continuum if the confinement is  
removed yielding an analytic function with a cut along the imaginary  
axis, that can be continued across the cut without singularities   
if no absorption of the  microsystem occurs. More  
precisely ${\cal T}(iy+\varepsilon)$ is considered  as practically  
constant for variations $\Delta y \approx {\hbar \over \tau_0}$, where   
$\tau_0$ has to be interpreted as a collision time.  
Treating expression (\ref{7}) we make use of the inequality  
        \begin{equation}  
        \label{9}  
        \left |  
        E_h - E_k  
        \right | \ll
        {  
        \hbar  
        \over  
             \tau_0  
        } ,  
        \end{equation}  
whose physical meaning is that the typical variation time $\tau_1$  
of the quantities $  
\langle  
{\hat A}  
\rangle_t  
$               is much larger than $\tau_0$.  
One then arrives at the  
following very perspicuous structure for  
$  
        {{{\cal U}^{'}(t)}}  
        \left(  
        {{{\hat a}^{\scriptscriptstyle \dagger}_{h}}{{\hat a}_{k}}}  
        \right)  
$:  
        \[  
        {{{\cal U}^{'}(t)}}  
        \left(  
        {{{\hat a}^{\scriptscriptstyle \dagger}_{h}}{{\hat a}_{k}}}  
        \right)  
        =  
        {{{\hat a}^{\scriptscriptstyle \dagger}_{h}}{{\hat a}_{k}}}  
        + t {\cal L}'
        \left(  
        {{{\hat a}^{\scriptscriptstyle \dagger}_{h}}{{\hat a}_{k}}}  
        \right)  
        \]  
        \begin{equation}  
        \label{tipstr}  
        {\cal L}'  
        \left(  
        {{{\hat a}^{\scriptscriptstyle \dagger}_{h}}{{\hat a}_{k}}}  
        \right)  
        =  
        {i\over\hbar}  
        \left[  
        {\hat H}_{\rm \scriptscriptstyle eff},  
        {\hat a}^{\scriptscriptstyle \dagger}_{h}  
        {\hat a}_{k}  
        \right]  
        - {1\over \hbar}  
        \left(  
        \left[  
        {\hat \Gamma}^{(1)} , {\hat a}^{\scriptscriptstyle\dagger}_h  
        \right]  
        {\hat a}_k  
        -  
        {\hat a}^{\scriptscriptstyle\dagger}_h  
        \left[  
        {\hat \Gamma}^{(1)}, {\hat a}_k  
        \right]  
        \right)  
        +  
        {1\over\hbar} \sum_\lambda  
        {\hat R}^{(1)}_{h \lambda}{}^{\dagger}  
        {\hat R}^{(1)}_{k \lambda},  
        \end{equation}  
where ${\hat H}_{\rm \scriptscriptstyle eff}  
        =  
        {\hat H}_0  
        +  
        {\hat V}^{\rm \scriptscriptstyle eff}$  
and  
        \begin{eqnarray*}  
        {\hat V}^{\rm \scriptscriptstyle eff}  
        \!\!&=&\!\!  
        \sum_{gr}  
        {\hat a}^{\scriptscriptstyle \dagger}_{r}  
        {\hat V}{}^{\rm \scriptscriptstyle eff}_{rg}
        {\hat a}_{g}  
        \\  
        \!\!&=&\!\!  
        i \hbar  
        \sum_{\lambda \lambda' \atop gr}  
        {\hat a}^{\scriptscriptstyle \dagger}_{r}  
        | \lambda \rangle \langle \lambda |  
        \frac 12  
        \left[  
        \left(  
        {\cal T}  
        \left({-{i \over \hbar} {E_r} +\varepsilon}\right)
         {\hat a}_r
        \right)  
        {{\hat a}^{\scriptscriptstyle \dagger}_{g}}
        +  
        {\hat a}_r  
        \left(  
        {\cal T}  
        \left({{i \over \hbar} {E_g} +\varepsilon}\right)  
         {\hat a}_g^{\scriptscriptstyle\dagger}  
        \right)  
        \right]  
        | \lambda' \rangle \langle \lambda' |  
        {\hat a}_{g}  
        \\  
        {\hat \Gamma}^{(1)}  
        \!\!&=&\!\!  
        \sum_{gr}  
        {\hat a}^{\scriptscriptstyle \dagger}_{r}  
        {\hat \Gamma}_{rg}  
        {\hat a}_{g}  
        \\  
        \!\!&=&\!\!  
         i \hbar  
        \sum_{\lambda \lambda' \atop gr}  
        {\hat a}^{\scriptscriptstyle \dagger}_{r}  
        | \lambda \rangle \langle \lambda |  
        \frac i2  
        \left[  
        \left(  
        {\cal T}  
        \left({-{i \over \hbar} {E_r} +\varepsilon}\right)
         {\hat a}_r
        \right)  
        {{\hat a}^{\scriptscriptstyle \dagger}_{g}}
        -  
        {\hat a}_r  
        \left(  
        {\cal T}  
        \left({{i \over \hbar} {E_g} +\varepsilon}\right)  
         {\hat a}_g^{\scriptscriptstyle\dagger}  
        \right)  
        \right]  
        | \lambda' \rangle \langle \lambda' |  
        {\hat a}_{g}  
        \\  
        {\hat R}^{(1)}_{k\lambda}  
        \!\!&=&\!\!  
        \sqrt{2\varepsilon \hbar^3}
        \sum_{g\lambda'}  
        {  
        \langle \lambda |  
        \left(  
        {\cal T}  
        \left({-{i \over \hbar} {E_g} +\varepsilon}\right)  
         {\hat a}_g^{\hphantom{\scriptscriptstyle \dagger}}  
        \right)  
        {{\hat a}^{\scriptscriptstyle \dagger}_{k}}  
        | \lambda' \rangle  
        \over  
        E_g + E_{\lambda} - E_k - E_{\lambda' } -  
        i\hbar\varepsilon  
        }  
        \langle \lambda' |  
         {\hat a}_g  
        \end{eqnarray*}  
and $| \lambda \rangle $ denotes an eigenvector of ${\hat H}_{\rm  
M}$ with eigenvalue $E_\lambda$ and of ${\hat H}_0$ with  
eigenvalue zero.  
Eq.(\ref{tipstr}) shows a typical structure arising in the  
calculation, which we will also find in the more complex  
situation examined in $\S\,3$, where the form of the  
different operators is further commented on.  
Let us observe that  
${\hat V}{}^{\rm \scriptscriptstyle
eff}_{rg}$  and ${\hat \Gamma}_{rg}$ are not c-number  
coefficients, but operators acting in the Fock-space for the  
macrosystem, as stressed by the hats; they are  
connected respectively  
to the self-adjoint and anti-self-adjoint part of  
what can be considered as an operator valued T-matrix.  
The last contribution displays the ``bilinear  
structure'' of the third term in the r.h.s. of (\ref{1}),  
connected to irreversibility and CP and not  
reproducible in the Hilbert space formalism, even resorting to an  
interaction potential which is not self-adjoint.  
Within the approximation leading to (\ref{tipstr}) one  
has ${\hat \Gamma}^{(1)} \approx {1\over2}  
         \sum_{h \lambda}  
        {\hat R}_{h \lambda}^{(1)}{}^{\dagger}  
        {\hat R}_{h \lambda}^{(1)}$ and
therefore  
${\cal L}^{'}{\hat N}=0$.  
Appealing to  
(\ref{alfa}) we may obtain an evolution equation  
for the matrix elements ${\varrho}_{fg}$ which is meaningful on a  
time scale much longer than the correlation time for ${\cal  
S}_{\rm M}$:  
        \[  
        {  
        d {\varrho}_{gf}          
        \over  
        dt  
        }  
        =  
        -{i \over \hbar}  
        \left(  
        {{E_g}-{E_f}}  
        \right)  
          {\varrho}_{gf}  
        +  
        {1 \over \hbar}  
        \sum_h  {\varrho}_{gh} {\mbox{\sf Q}}^{\scriptscriptstyle  
         \dagger}_{hf}  
        +  
        {1 \over \hbar}  
        \sum_k  {\mbox{\sf Q}}_{gk} {\varrho}_{kf}  
        +  
        {1 \over \hbar}  
        \sum_{hk \atop \lambda\xi}  
        \left(  
        {{\mbox{\sf L}}_{\lambda\xi}}  
        \right)_{gk}  
        {\varrho}_{kh}  
        \left(  
        {{{\mbox{\sf L}}_{\lambda\xi}}}  
        \right)^*_{fh}  
        \]  
with  
        \begin{eqnarray*}  
        {\mbox{\sf Q}}_{kf}  
        &=&  
        \hbar{\hbox{\rm Tr}}_{{{{\cal H}}}}  
        \left[{  
        \left(  
        {\cal T}  
        \left({-{i \over \hbar} {E_k} +\varepsilon}\right)  
           {{\hat a}_{k}}  
        \right)  
        {{\hat a}^{\scriptscriptstyle \dagger}_{f}}  
        {{\hat \varrho}_{\rm M}(t)}  
        }\right]  
        \\  
        \left(  
        {{\mbox{\sf L}}_{\lambda\xi}}  
        \right)  
        _{kf}  
        &=&  
        \sqrt{2\varepsilon\hbar^3 \pi_\xi}  
        {\langle  
        \lambda  
        \vert  
        \left[{  
        \left(  
        {{\cal T}  
        \left({-{i \over \hbar} {E_k} +\varepsilon}\right)  
        {{\hat a}_{k}}}  
        \right)  
        {{\hat a}^{\scriptscriptstyle \dagger}_{f}}  
        }\right]  
        {  
         1  
        \over  
        {{E_k}+{E_{{\lambda}}}-{E_f}-{H}_{\rm M}  
        -i\hbar\varepsilon}  
        }  
        \vert  
        {\xi(t)}  
        \rangle}                               ;  
        \end{eqnarray*}  
$\xi(t)$ is a complete system of  
eigenvectors of  
${{\hat \varrho}_{\rm M}(t)}$,  
$({{\hat \varrho}_{\rm M}(t)}=\sum_{\xi(t)}  
\pi_{\xi(t)}  
\vert {{\xi(t)}} \rangle  
{\langle \xi(t) \vert})$.  
To show the connection with (\ref{1}) we introduce  
in ${{\cal H}^{(1)}}$   the operators  ${\hat {\mbox{\sf Q}}}^{(1)},  
{\hat {\mbox{\sf L}}}^{(1)}_{\lambda\xi}$:  
        \[  
        \langle  
        {k}  
        \vert  
        {\hat {\mbox{\sf Q}}}^{(1)}  
        \vert  
        {f}  
        \rangle  
        =\mbox{\sf Q}_{kf}  
        \quad  
        ,  
        \quad  
        \langle  
        {k}  
        \vert  
        {{\hat {\mbox{\sf L}}}^{(1)}_{\lambda\xi}}  
        \vert  
        {f}  
        \rangle  
        =  
         {\bigl( {{\mbox{\sf L}}_{\lambda\xi}} \bigr)}_{kf} ,  
        \]  
thus attaining in the Schr\"odinger picture the full evolution of  
${\varrho}^{(1)}$, given by the typical  Lindblad generator:  
        \begin{equation}  
        {  
        d {\hat {\varrho}}^{(1)}  
        \over  
                      dt  
        }  
        =  
        -{i \over \hbar}  
        \left[  
	{  
        {\hat {\mbox{\sf H}}}_{\rm eff},{\hat {\varrho}}^{(1)}}  
	\right]  
        +  {1\over 2\hbar}  
        \left \{  
        {  
        \left(  
        {\hat {\mbox{\sf Q}}}^{(1)}+  
        {{\hat {\mbox{\sf Q}}}^{(1)}}{}^{ \dagger}  
        \right)  
        ,{\hat {\varrho}}^{(1)}}  
                \right \}  
        +  
        {1 \over \hbar}  
        \sum^{}_{{\xi,\lambda  }}  
        {\hat {\mbox{\sf L}}}^{(1)}_{\lambda\xi} {\hat {\varrho}}^{(1)}  
        {\hat {\mbox{\sf L}}}^{(1)}_{\lambda\xi}{}^{  
        \dagger}\ ,  
        \label{Meq}  
        \end{equation}  
where  
       ${{\hat {\mbox{\sf H}}}}^{(1)}_{\rm \scriptscriptstyle eff}  
       ={\hat {\mbox{\sf H}}}^{(1)}_0 + {i\over 2}  
        \left(  
        {{\hat {\mbox{\sf Q}}}^{(1)}-  
        {{\hat {\mbox{\sf Q}}}^{(1)}}{}^{ \dagger}}  
        \right)$.  
Furthermore according to preservation of trace we have  
        ${{\hat {\mbox{\sf Q}}}^{(1)}+  
        {{\hat {\mbox{\sf Q}}}^{(1)}}{}^{ \dagger}}  
        =  
        -  
        \sum_{\lambda\xi}  
        {\hat {\mbox{\sf L}}}^{(1)}_{\lambda\xi}{}^{ \dagger}  
        {\hat {\mbox{\sf L}}}^{(1)}_{\lambda\xi}$.  
\par 
As it is well-known (\ref{Meq}) is  
apt to describe very different physical situations. If  
the last contribution, which we will call ``incoherent'', may be  
neglected, at least as a first approximation, eq.(\ref{Meq}) is  
equivalent to a Schr\"odinger equation  with a possibly complex  
potential. In the case of a particle interacting with matter this  
equation is well-suited to describe a coherent optical behaviour,  
for example in terms of a refractive index, as it is usually done  
in neutron optics~\cite{Erice,Sears} and recently also in atom  
optics~\cite{Mlynek,Vigue}. In this framework the operator  
${\hat {\mbox{\sf Q}}}^{(1)}$  
is to be interpreted as an optical potential, which in our  
formalism is naturally linked to matrix elements of the  
T-operator, thus showing the connection between the effective,  
macroscopic description through an index of refraction and  
quantities characterising the local interactions. The  
T-operator  
may be replaced by phenomenological expressions (for  
example the Fermi pseudo-potential  
in the case of neutron optics).  
This picture is particularly useful to deal with particle  
interferometry. To see how the last contribution may be linked to  
an interaction having a measuring character let us introduce  
the reversible mappings  
$  
{\cal A}_{t'' t'}={\hat U}^{(1)}_{t'' t'} \cdot  
{\hat U}^{(1)}_{t'' t'}{}^{\dagger}  
$, where ${\hat U}_{t'' t'}^{(1)}  
        = T  
        \exp            (  
        {-{i \over \hbar} \int_{t'}^{t''} dt \,  
        ({{\hat {\mbox{\sf H}}}^{(1)}_0(t)+i{\hat {\mbox{\sf Q}}}^{(1)}(t)}  
        }                )  
        )  
        $,  
corresponding to a coherent contractive evolution of the microsystem during  
the time interval $[t',t'']$, and  
the CP mappings  
       ${\cal L}_{\lambda \xi}=  
        {\hat {\mbox{\sf L}}}_{\lambda\xi}^{(1)}(t) \cdot  
        {\hat {\mbox{\sf L}}}^{(1)}_{\lambda\xi}{}^{  
        \dagger}(t)$,  
having a measuring character, as it is clear from their very  
structure, reminiscent of the reduction postulate.  
The expression of the operators   ${\hat {\mbox{\sf L}}}^{(1)}_{\lambda\xi}$  
shows how these mappings may be linked with a transition  
inside the macrosystem  
 specified by the pair of indexes  
$\xi,\lambda$,  
as a result of scattering with the microsystem.  
These transitions are in general not detectable,  
but under suitable conditions they could prime real events.  
The solution of  (\ref{Meq}) may be  
written in the form:  
        \begin{equation}  
        \label{Subcoll}  
        {\hat {\varrho}}_t  
        =  
        {\cal A}_{t t_0}{\hat {\varrho}}_{t_0} +  
        \sum_{\lambda_1 \xi_1}  
        \int_{t_0}^{t} dt_1 \, {\cal A}_{t t_1} {\cal L}_{\lambda_1  
        \xi_1}(t_1)  
        {\cal A}_{t_1 t_0}{\hat {\varrho}}_{t_0} +{}  
        \ldots \, ,  
        \end{equation}  
that is a sum over subcollections corresponding to the  
realization of no event, one event and so on.  
The set of variables $N_{\lambda\xi}(\tau)$, $\tau\geq t_0$,  
(number of transitions  
up to  
time $\tau$), define a multicomponent classical stochastic  
process, and (\ref{Subcoll}) corresponds to the decomposition of  
the evolution map on the space of trajectories for  
$N_{\lambda\xi}(\tau)$.  
This is a straightforward generalization of the typical  
``counting process'' considered by Srinivas and  
Davies~\cite{Srinivas}. Of course other decompositions in terms  
of operation valued maps are possible on trajectory spaces  
related to different observables, as indicated in  
$\S\,1$. When the first term in (\ref{Subcoll}) is largely  
predominant a wavelike description as given by the Schr\"odinger  
equation is sufficiently accurate and small disturbances,  
conveyed by the other terms, play no significant role.  
In a different physical context however, as would be the case for  
the brownian motion of a particle interacting with an ideal gas,  
the interplay between the contractive and the incoherent part has  
a major role.  
Being  
interested in the dynamics far away from the walls the quantum  
number $h$ corresponds to the momentum variable ${\mbox{\bf  
p}_h}$, and supposing that momentum transfers are small  
(Fokker-Planck approximation) one arrives at an equation  
describing diffusion in phase-space which has the following form  
(see~\cite{Diosi}):  
        \[  
        {  
        d {\hat \varrho}  
        \over  
                dt  
        }  
        =  
        -{i\over\hbar}  
        \left[  
        {{\hat {\mbox{\sf H}}}_{\rm \scriptscriptstyle eff}} ,{\hat \varrho}  
        \right]  
        -  
        D_{pp}  
        \left[  
        {\hat {\mbox{\sf x}}},  
        \left[  
        {\hat {\mbox{\sf x}}},{\hat \varrho}  
        \right]  
        \right]  
        -  
        D_{qq}  
        \left[  
        {\hat {\mbox{\sf p}}},  
        \left[  
        {\hat {\mbox{\sf p}}},{\hat \varrho}  
        \right]  
        \right]  
        -{  
        i\eta  
        \over  
             2 M  
        }  
        \left[  
        {\hat {\mbox{\sf x}}} ,  
        \left \{  
        {\hat {\mbox{\sf p}}},{\hat \varrho}  
        \right \}  
        \right]               ,  
        \]  
$D_{qq},  
D_{qq}$ and $\eta$ being diffusion coefficients linked in  
different ways to the operators ${\hat {\mbox{\sf L}}}_{\lambda\xi}^{(1)}$  
and $M$ the mass of the particle.  
\par  
\section{Theory of a  macrosystem: thermodynamic evolution by a  
scattering map}  
\par  
\setcounter{equation}{0}  
We consider a very schematic model of  macrosystem in the  
non-relativistic case built by one type of molecules with mass $m$  
confined inside a region $\omega$, interacting by a two body  
potential $V(  
\left |  
{\mbox{\bf x}} -{\mbox{\bf y}}  
\right |  
)$;  
for the sake of simplicity no internal structure of the  
molecules is taken into account. In the field theoretical  
language  the system is described by a quantum  
Schr\"odinger field (QSF):  
        \begin{equation}  
        \label{campo}  
        {\hat \psi}({\mbox{\bf x}}) = \sum_f u_f ({\mbox{\bf x}}) {\hat  
        a}_f  
        \qquad  
        {  
        \left[  
        {\hat a}_f , {\hat a}^{\scriptscriptstyle\dagger}_g  
        \right]  
        }_\pm = \delta_{fg}  
        \end{equation}  
        \[  
        -{  
        \hbar^2  
        \over  
        2m  
        }  
        \Delta_2  
        u_f({\mbox{\bf x}})= E_f  
        u_f({\mbox{\bf x}}),  
        \qquad  
        u_f({\mbox{\bf x}})=0 \quad {\mbox{\bf x}}\in  
        \partial\omega  .  
        \]  
We shall assume the following  
Hamiltonian to take local interactions and confinement  
into account:  
        \begin{equation}  
        \label{10}  
        {\hat H}=  
        \sum_f E_f  
        {\hat a}_f^{\scriptscriptstyle\dagger} {\hat a}_f  
        + {1\over 2}  
        \sum_{l_1 l_2 \atop f_1 f_2}  
        {\hat a}_{l_1}^{\scriptscriptstyle\dagger}  
        {\hat a}_{l_2}^{\scriptscriptstyle\dagger}  
        V_{l_1 l_2 f_2 f_1}  
        {\hat a}_{f_2}  
        {\hat a}_{f_1}  
        \end{equation}  
        \begin{equation}  
        \label{11}  
        V_{l_1 l_2 f_2 f_1}  
        =  
        \int_{\omega} d^3 \! {\mbox{\bf x}}  
        \int_{\omega} d^3 \! {\mbox{\bf y}}  
        \,  
        u_{l_1}^{*}({\mbox{\bf x}})  
        u_{l_2}^{*}({\mbox{\bf y}})  
        V(  
        \left |  
        {\mbox{\bf x}} -{\mbox{\bf y}}  
        \right |  
        )      
	u_{f_2}({\mbox{\bf y}})  
        u_{f_1}({\mbox{\bf x}})  
              .  
        \end{equation}  
Eq.(\ref{campo}) is linked to the basic ``local''  
Hamiltonian for the non-confined field (NC)  
        \begin{eqnarray*}  
        {\hat H}_{\rm \scriptscriptstyle NC}  
        &=&  
        \int d^3\! {\mbox{\bf  x}}  
        \,  
        {{  
        \hbar^2  
        \over  
               2m  
        }}  
        {\mbox{\rm grad}}  
        {\hat \psi}_{\rm \scriptscriptstyle NC}^{\scriptscriptstyle\dagger}
	({\mbox{\bf x}})  
        \cdot  
        {\mbox{\rm grad}}  
        {\hat \psi}_{\rm \scriptscriptstyle NC}({\mbox{\bf x}})  
        \\  
        &\hphantom{=}&  
        +  
        {1\over 2}  
        \int d^3\! {\mbox{\bf  x}}  
        d^3\! {\mbox{\bf  r}}  
	\,  
        {\hat \psi}_{\rm \scriptscriptstyle NC}^{\scriptscriptstyle\dagger}  
        \left(  
        {\mbox{\bf x}}-  
        {{\mbox{\bf r}}\over 2}  
        \right)  
                {\hat \psi}_{\rm \scriptscriptstyle NC}^{\scriptscriptstyle
	\dagger}  
                \left(  
                {\mbox{\bf x}}+  
        {{\mbox{\bf r}}\over 2}  
                \right)  
                        V(r)  
        {\hat \psi}_{\rm \scriptscriptstyle NC}  
        \left(  
        {\mbox{\bf x}}+{{\mbox{\bf r}}\over 2}  
        \right)  
        {\hat \psi}_{\rm \scriptscriptstyle NC}  
        \left(  
        {\mbox{\bf x}}-{{\mbox{\bf r}}\over 2}  
        \right)  
        \end{eqnarray*}  
        \[  
        {  
        \left[  
        {\hat \psi}_{\rm \scriptscriptstyle NC}({\mbox{\bf x}}),  
        {\hat \psi}_{\rm \scriptscriptstyle  
        NC}^{\scriptscriptstyle\dagger}({\mbox{\bf x}}')  
        \right]  
        }_\pm  
        =  
        \delta ({\mbox{\bf x}}-{\mbox{\bf x}}'),  
        \]  
simply selecting the part of ${\hat H}_{\rm \scriptscriptstyle NC}$ related 
to the  
``normal modes'' $u_f$ typical of the confinement: in fact the  
preparation procedure should imply a kind of relaxation of  
${\hat \psi}_{\rm \scriptscriptstyle NC}({\mbox{\bf x}})$  
to ${\hat {\psi}}({\mbox{\bf x}})$.  
Skipping this problem  
and also the related question of the full explicit structure of  
${\cal L}^{'}$ for a realistic system, we shall  
simply take the Hamiltonian (\ref{10}) containing only the normal  
modes of the field inside $\omega$.  
If we are interested in a hydrodynamic description relevant  
observables  are constructed starting with the densities of the  
typical constants of motion, mass and energy:  
        \begin{eqnarray}  
        \label{12}  
        {\hat \rho}_{m}({\mbox{\bf x}})
        &=&  
        m  
        {\hat \psi}^{\scriptscriptstyle\dagger}({\mbox{\bf x}})  
        {\hat \psi}({\mbox{\bf x}})  
        \\  
        {\hat e}({\mbox{\bf x}})  
        &=&  
        {{  
        \hbar^2  
        \over  
               2m  
        }}  
        {\mbox{\rm grad}}  
        {\hat \psi}^{\scriptscriptstyle\dagger}({\mbox{\bf x}})  
        \cdot  
        {\mbox{\rm grad}}  
        {\hat \psi}({\mbox{\bf x}})  
        \nonumber \\  
        &\hphantom{=}&  
        +  
        {1\over 2}  
        \int_{\omega_x} d^3\! {\mbox{\bf  r}}  
	\,  
        {\hat \psi}^{\scriptscriptstyle\dagger}  
        \left(  
        {\mbox{\bf x}}-  
        {{\mbox{\bf r}}\over 2}  
        \right)  
                {\hat \psi}^{\scriptscriptstyle\dagger}  
                \left(  
                {\mbox{\bf x}}+  
        {{\mbox{\bf r}}\over 2}  
                \right)  
                        V(r)  
        {\hat \psi}  
        \left(  
        {\mbox{\bf x}}+{{\mbox{\bf r}}\over 2}  
        \right)  
        {\hat \psi}  
        \left(  
        {\mbox{\bf x}}-{{\mbox{\bf r}}\over 2}  
        \right)  
         ,  
        \nonumber  
        \end{eqnarray}  
where the dependence of $\omega_{x}$ on ${\mbox{\bf x}}$  
is generally negligible if $V(r)$ is a short range  
potential.  
In the case of a kinetic description we replace (\ref{12}) by the  
``Boltzmann'' operator density  
${\hat f}({\mbox{\bf x}},{\mbox{\bf p}}) =  
        m \sum_{hk}  
        {\hat a}_h^{\scriptscriptstyle\dagger}  
        \langle  
        u_h |   {\hat {\mbox{\sf F}}}^{(1)}   
        ({\mbox{\bf x}},{\mbox{\bf p}})  | u_k         
        \rangle  
        {\hat a}_k$,  
where ${\hat {\mbox{\sf F}}}^{(1)}$ is the density of  joint one  
particle position-momentum observables~\cite{Lanz5,Holevo}.  
These densities lead to slowly varying quantities if they  
are integrated over regions large enough in space or phase-space,  
since one has constants if the integration is extended  
over the whole space. The constants of motion leading to this  
subdynamics are linked to very fundamental  
symmetries: time translation invariance and gauge symmetry.  
Our relevant observables have the general structure:  
        \begin{equation}  
        \label{relobs}  
        \sum_{hk}  
        {\hat a}_h^{\scriptscriptstyle\dagger}  
        A_{hk}(\xi)  {\hat a}_k  
        \quad ,  
        \quad  
        \sum_{k_1 k_2 \atop h_1 h_2}  
        {\hat a}_{h_1}^{\scriptscriptstyle\dagger}  
        {\hat a}_{h_2}^{\scriptscriptstyle\dagger}  
        A_{h_1 h_2 k_2 k_1} ({\mbox{\bf x}})  
        {\hat a}_{k_2}  
        {\hat a}_{k_1}  
        \end{equation}  
        \[  
        A_{h_1 h_2 k_2 k_1} ({\mbox{\bf x}})  
        = \frac 12  
        \int_{\omega_x} d^3\! {\mbox{\bf r}} \,  
        u_{h_1}^{*}  
        \left(  
        {\mbox{\bf x}}-  
        {{\mbox{\bf r}}\over 2}  
        \right)  
        u_{h_2}^{*}  
        \left(  
        {\mbox{\bf x}}+ {{\mbox{\bf r}}\over 2}  
        \right)  
        V(r)  
        u_{k_2}  
        \left(  
        {\mbox{\bf x}}+{{\mbox{\bf r}}\over 2}  
        \right)  
        u_{k_1}  
        \left(  
        {\mbox{\bf x}}-{{\mbox{\bf r}}\over 2}  
        \right)  
        .  
        \]  
We thus have to study in Heisenberg picture the  
expressions:  
        \begin{equation}  
        \label{37}  
        \sum_{hk}  
        e^{{i\over \hbar}{\hat H}t}  
        {\hat a}_h^{\scriptscriptstyle\dagger}  
        {\hat a}_k  
        e^{-{i\over \hbar}{\hat H}t}  
        A_{hk}(\xi) ,  
        \qquad  
        \sum_{h_1 h_2 \atop k_1 k_2}  
        e^{{i\over \hbar}{\hat H}t}  
        {\hat a}_{h_1}^{\scriptscriptstyle\dagger}  
        {\hat a}_{h_2}^{\scriptscriptstyle\dagger}  
        {\hat a}_{k_2}  
        {\hat a}_{k_1}  
        e^{-{i\over \hbar}{\hat H}t}  
        A_{h_1 h_2 k_2 k_1} ({\mbox{\bf x}}) ,  
        \end{equation}  
and shall take into account that by the slow variability  
only terms that are ``diagonal enough'' are really relevant;  
the sums should be restricted to  indexes such that:  
        \[  
        \frac 1{\hbar} |E_h - E_k| < {1\over\tau_1} \qquad  
        \frac 1{\hbar} |E_{h_1} +E_{h_2}- E_{k_1}-E_{k_2}| <  
        {1\over\tau_1},  
        \]  
where $\tau_1$ is the characteristic variation time of the  
relevant quantities.  
We would like to stress the fact that  the QSF   
is the basic tool to describe a massive continuum, just  
like the quantum electromagnetic field describes a massless continuum.  
The  dynamics of the QSF,  
$  
{\hat \psi}({\mbox{\bf x}},t)=  
        e^{{i\over \hbar}{\hat H}t}  
{\hat \psi}({\mbox{\bf x}})  
        e^{-{i\over \hbar}{\hat H}t}  
$, in terms of which one can rewrite (\ref{37}), is given by the  
simple field equation  
        \begin{equation}  
        \label{13}  
        i\hbar {\partial\over  \partial t}  
        {\hat \psi}({\mbox{\bf x}},t)  
        =-{  
          \hbar^2  
        \over  
                 2m  
        }          \Delta_2  
        {\hat \psi}({\mbox{\bf x}},t)  
        +  
        \int d^3\! {\mbox{\bf y}}  
        \,  
        {\hat \psi}^{\scriptscriptstyle\dagger}({\mbox{\bf y}},t)  
        V(  
        \left |  
        {\mbox{\bf x}}- {\mbox{\bf y}}  
        \right |  
        )  
        {\hat \psi}({\mbox{\bf y}},t)  
        {\hat \psi}({\mbox{\bf x}},t) ,  
        \end{equation}  
however no such equation holds  
for the expectation value of the field  
        $        {\psi}({\mbox{\bf x}},t)=  
        \langle  
        {\hat \psi}({\mbox{\bf x}},t)  
        \rangle $  
due to correlations in the non-linear term;  
${\psi}({\mbox{\bf x}},t)$ is not useful to calculate the  
expectations of operators (\ref{37}) and  
therefore a classical Schr\" odinger field equation for  
${\psi}({\mbox{\bf x}},t)$ has no physical meaning in general. In  
this respect the case of electromagnetism, where no  
self-interaction of the field occurs, is deeply different and  
allows classical electrodynamics to play an important role. To the  
macrosystem one associates typical ``thermodynamic state''  
parameters: the velocity field ${{\mbox{\bf v}}}({\mbox{\bf x}},t)$, the  
temperature field $\beta({\mbox{\bf x}},t)$, the chemical  
potential field $\mu ({\mbox{\bf x}},t)$ in the case of the  
hydrodynamic description or more generally a field  
$\mu({\mbox{\bf x}},{\mbox{\bf p}},t)$ on the one-particle  
phase-space in the kinetic case~\cite{Roepke}. The parameters  
$\beta({\mbox{\bf x}},t)$ and $\mu ({\mbox{\bf x}},t)$  
($\mu({\mbox{\bf x}},{\mbox{\bf p}},t)$) determine the  
expectation values of energy density and mass density (the  
Boltzmann operator); let us briefly recall how the relation  
between state variables and expectation values is  
established~\cite{Robin}.
At any time $t$ one considers the whole set of  statistical
operators $  
\left \{  
{\hat w}  
\right \}  
$  
which yield  the expectation values assigned at that time:
        \[  
        \langle  
        {\hat e}^{(0)}({\mbox{\bf x}})
        \rangle_t
        \! = \!  
        {\mbox{\rm Tr}} \!  
        \left(  
        {\hat e}^{(0)}({\mbox{\bf x}}) {\hat w}  
        \right)  
         , \
        \langle  
        {\hat \rho}_{m}({\mbox{\bf x}})
        \rangle_t
        \! = \!  
        {\mbox{\rm Tr}}   \!  
        \left(  
        {\hat \rho}_{m}({\mbox{\bf x}}) {\hat w}
        \right)  
        ,    \
        \langle  
        {\hat f}^{(0)}({\mbox{\bf x}},{\mbox{\bf p}})  
        \rangle_t
        \! = \!  
        {\mbox{\rm Tr}}     \!  
        \left(  
        {\hat f}^{(0)}({\mbox{\bf x}},{\mbox{\bf p}})  
        {\hat w}  
        \right)  
        \]  
        \begin{eqnarray}  
        \label{densita}  
        {\hat e}^{(0)}({\mbox{\bf x}})  
        &=&  
        \frac {1}{2m}  
        \left(  
        i\hbar {\partial\over\partial{\mbox{\bf x}}} - m{{\mbox{\bf v}}}  
	({\mbox{\bf x}},t)  
        \right)  
        {\hat \psi}^{\scriptscriptstyle\dagger}({\mbox{\bf x}})  
        \cdot  
        \left(  
        -i\hbar {\partial\over\partial{\mbox{\bf x}}} -  
        m{{\mbox{\bf v}}}({\mbox{\bf x}},t)  
        \right)  
        {\hat \psi}({\mbox{\bf x}})  
        \\  
        &\hphantom{=}&  
        +  
        \frac 12  
        \int_{\omega_x} d^3\! {\mbox{\bf r}} \,  
        {\hat \psi}^{\scriptscriptstyle\dagger}  
        \left(  
        {\mbox{\bf x}}-  
        {{\mbox{\bf r}}\over 2}  
        \right)  
                {\hat \psi}^{\scriptscriptstyle\dagger}  
                \left(  
                {\mbox{\bf x}}+  
        {{\mbox{\bf r}}\over 2}  
                \right)  
                        V(r)  
        {\hat \psi}  
        \left(  
        {\mbox{\bf x}}+{{\mbox{\bf r}}\over 2}  
        \right)  
        {\hat \psi}  
        \left(  
        {\mbox{\bf x}}-{{\mbox{\bf r}}\over 2}  
        \right)  
        \nonumber  
        \\  
        {\hat \rho}_{m}({\mbox{\bf x}})
        &=&  
        {\hat \rho}^{(0)}_{m}({\mbox{\bf x}})
        =  
        m  
        {\hat \psi}^{\scriptscriptstyle\dagger}({\mbox{\bf x}})  
        {\hat \psi}({\mbox{\bf x}})  
        ,             \qquad  
        {\hat f}^{(0)}({\mbox{\bf x}},{\mbox{\bf p}})  
        =  
        {\hat f}({\mbox{\bf x}},{\mbox{\bf p}}-m{\mbox{\bf  
        v}}({\mbox{\bf x}},t))  
        \nonumber  
        \end{eqnarray}  
where
the quantities indexed by $(0)$
(depending   
explicitly on the velocity field ${{\mbox{\bf v}}}  
({\mbox{\bf x}},t)$) represent densities in
the reference frame in which the continuum is locally at rest.  
The velocity field is related to the expectation value of the  
momentum density ${\hat {\mbox{\bf p}}}({\mbox{\bf x}})$ through the relation  
$  
\langle  
{\hat {\mbox{\bf p}}}^{(0)}({\mbox{\bf x}})  
\rangle =0  
$, where  
        \[  
        {\hat {\mbox{\bf p}}}^{(0)}({\mbox{\bf x}})  
        =  
        \frac {1}{2}     \!  
        \left \{  
        {\hat \psi}^{\scriptscriptstyle\dagger}({\mbox{\bf x}})  
        \left(  
        -i\hbar {\partial\over\partial{\mbox{\bf x}}} - m{{\mbox{\bf v}}}
	({\mbox{\bf x}},t)  
        \right)  
        {\hat \psi}({\mbox{\bf x}})  
        +  
        \left[  
        \left(  
        i\hbar {\partial\over\partial{\mbox{\bf x}}} - m{{\mbox{\bf v}}}
	({\mbox{\bf x}},t)  
        \right)  
        {\hat \psi}^{\scriptscriptstyle\dagger}({\mbox{\bf x}})  
        \right]  
        {\hat \psi}({\mbox{\bf x}})  
        \right \}  
        \]  
or equivalently  $      \langle  
        {\hat {\mbox{\bf p}}}({\mbox{\bf x}})  
        \rangle_t =
        {{\mbox{\bf v}}}({\mbox{\bf x}},t)  
        \langle  
        {\hat \rho}_{m}({\mbox{\bf x}})
        \rangle_t   $.
Then one looks for a  statistical operator in the set $  
\left \{  
{\hat w}  
\right \}  
$      such that the Von-Neumann entropy $S=-k{\mbox{\rm Tr}  
\left(  
{\hat w} \log {\hat w}  
\right)  
}$                   is maximal, i.e. the most unbiased choice of  
a  statistical operator leading to the given expectation values.  
The unique solution of this problem is  
        \begin{equation}  
        \label{14}  
        {\hat w}[\beta(t) , \mu (t), {{\mbox{\bf v}}}(t)]  
        =  
        {  
        e^{-{  
        \int_{\omega} d^3\! {\bf \scriptscriptstyle x} \,  
        \beta({\bf \scriptscriptstyle x},t)  
        \left[  
        {\hat e}^{(0)}({\bf \scriptscriptstyle x})  
        -  
        \mu({\bf \scriptscriptstyle x},t){\hat \rho}_{m}
	({\bf \scriptscriptstyle x})
        \right]  
        }}  
        \over  
        {\mbox{{\rm Tr}}} \, 
        e^{-{  
        \int_{\omega} d^3\! {\bf \scriptscriptstyle x} \,  
        \beta({\bf \scriptscriptstyle x},t)  
        \left[  
        {\hat e}^{(0)}({\bf \scriptscriptstyle x})  
        -  
        \mu({\bf \scriptscriptstyle x},t){\hat \rho}_{m}
	({\bf \scriptscriptstyle x})
        \right]  
        }}  
        }  
        \end{equation}  
and analogously in the kinetic case.  
\par \noindent 
The corresponding 
$S=-k{\mbox{\rm Tr}  
(  
{\hat w}[\beta(t) , \mu (t), {{\mbox{\bf v}}}(t)]  
\log {\hat w}[\beta(t) , \mu (t), {{\mbox{\bf v}}}(t)]  
)  
}$  
is the  thermodynamic entropy of the  macrosystem.  
If the time evolution of the expectation  
values  
$  
        \langle  
        {\hat e}^{(0)}({\mbox{\bf x}})  
        \rangle_t  
        ,  
        \langle  
        {\hat \rho}_{m}({\mbox{\bf x}})
        \rangle_t  
        ,  
        \left(  
        \langle  
        {\hat f}({\mbox{\bf x}},{\mbox{\bf p}})  
        \rangle_t  
        \right)  
$  
is given by the Hamiltonian evolution (\ref{37})  
or more generally by a map  
${\cal M}^{'}_{t t_0}$,  having a preadjoint  
${\cal M}_{t t_0}$  which does not decrease the Von-Neumann  
entropy,  
one immediately has that the thermodynamic entropy  
is non-decreasing. In this way one establishes the second  
principle of thermodynamics on a very clear dynamical basis.  
\par      
In the simplest scheme of macroscopic dynamics the thermodynamic  
state parameters ${{\mbox{\bf v}}}({\mbox{\bf x}},t)$,
$\beta({\mbox{\bf x}},t)$, $\mu ({\mbox{\bf x}},t)$ 
($\mu({\mbox{\bf x}},{\mbox{\bf p}},t)$) at time $t_0$ determine  
its evolution for $t>t_0$, e.g., by differential equations.
Phenomenology shows that this is very often the case.  
Tackling the problem from the theoretical viewpoint one is  
induced, considering the operators  
	\[
        \begin{array}{ccccccc}  
        {\dot  {\hat \rho}}_{m}({\mbox{\bf x}})
        &=&  
        {i\over\hbar}[{\hat H},{{\hat \rho}}_{m}({\mbox{\bf x}})]
        &\qquad&  
        {\dot  {\hat { \mbox{\bf p}}}}({\mbox{\bf x}})  
        &=&  
        {i\over\hbar}[{\hat H},{{\hat {\mbox{\bf p}}}}({\mbox{\bf x}})]  
        \\  
        {\dot {\hat e}}({\mbox{\bf x}})  
        &=&  
        {i\over\hbar}[{\hat H},{{\hat e}}({\mbox{\bf x}})]  
        &\qquad&  
        (  
        {\dot {\hat f}}({\mbox{\bf x}},{\mbox{\bf p}})  
        &=&  
        {i\over\hbar}[{\hat H},{{\hat f}}({\mbox{\bf  
        x}},{\mbox{\bf p}})]  
        )       ,
        \end{array}
	\]  
to calculate their expectations with the  statistical operator  
given by (\ref{14}). This leads to wrong results as can be seen  
from the fact that the expectation values of the currents which  
can be associated, through a conservation equation, to these  
operators would vanish~\cite{Zubarev}, due to time reversal  
invariance of microphysics, thus failing to describe any  
dissipative flow (e.g., heat conduction, viscosity, etc.). The
idea of a time scale for the thermodynamic evolution and of a  
related subdynamics for the basic densities leads to a refinement  
of the aforementioned procedure: assume that  
${i\over\hbar}[{\hat H},\cdot]$ can be replaced by a mapping  
${\cal L}^{'}$, initially defined on the linearly independent  
elements  
$  
{\hat a}^{\scriptscriptstyle \dagger}_{h}  
{\hat a}_{k}  
$,  
$  
{\hat a}^{\scriptscriptstyle \dagger}_{h_1}  
{\hat a}^{\scriptscriptstyle \dagger}_{h_2}  
{\hat a}_{k_2}
{\hat a}_{k_1}
$,
giving the slow time evolution of the relevant variables. In  
this way not only the  statistical operator  
${\hat w}[\beta(t) , \mu (t), {{\mbox{\bf v}}}(t)]  
$, but also the evolution operator is tuned to the relevant  
observables. Then one has the following set of closed evolution  
equations for the thermodynamic fields  
${{\mbox{\bf v}}}({\mbox{\bf x}},t)$,  
$\beta({\mbox{\bf x}},t)$, $\mu ({\mbox{\bf x}},t)$,  
($\mu({\mbox{\bf x}},{\mbox{\bf p}},t)$)  
related to the basic observables  
$  
{\hat A}={  {\hat \rho}}_{m}({\mbox{\bf x}}),\,
{  {\hat {\mbox{\bf p}}}}({\mbox{\bf x}}),\,{{\hat e}}({\mbox{\bf x}}),\,  
({ {\hat f}}({\mbox{\bf x}},{\mbox{\bf p}}))  
$:  
        \begin{equation}  
        \label{one}  
        {  
        d  
        \over  
         dt  
        }  
        {\mbox{\rm Tr}}  
        \left(  
        {\hat A}  
        {\hat w}[\beta(t) , \mu (t), {{\mbox{\bf v}}}(t)]  
        \right)  
        =  
        {\mbox{\rm Tr}}  
        \left(  
        ({\cal L}^{'}  
        {\hat A})  
        {\hat w}[\beta(t) , \mu (t), {{\mbox{\bf v}}}(t)]  
        \right)  .  
        \end{equation}  
The non-Hamiltonian form of the map ${\cal L}^{'}$ eliminates the  
aforementioned difficulties with vanishing dissipative flows;  
preliminary investigations of the consequences of (\ref{one}) in  
the case of a dilute gas indicate that it could be the right  
solution.  
The map ${\cal L}^{'}$ that adequately replaces  
${i\over\hbar}[{\hat H},\cdot]$ for the slow variables must  
generate an evolution of the relevant observables that preserves  
their positivity properties (e.g.,
${  {\hat \rho}}_{m}({\mbox{\bf x}}),
{ {\hat f}}({\mbox{\bf x}},{\mbox{\bf p}})  
$) and also conservation of mass (${\hat M}=  
\int d^3 \! {\mbox{\bf x}} \,  
{  {\hat \rho}}_{m}({\mbox{\bf x}})=
\int d^3 \! {\mbox{\bf x}}d^3 \! {\mbox{\bf p}}\,  
{ {\hat f}}({\mbox{\bf x}},{\mbox{\bf p}})  
$) and of energy (${\hat E}=  
\int d^3 \! {\mbox{\bf x}} \,  
 {  {\hat e}}({\mbox{\bf x}})  
$). Then  
${\cal L}^{'}{\hat M}=0$ and  
${\cal L}^{'}{\hat E}=0$, while positivity with respect to  
observables constructed in terms of creation and annihilation  
operators could arise by a stronger property, reminding CP:  
        \begin{equation}  
        \label{a}  
        \sum_{hk}  
        \langle  
        \psi_h  
        |  
        {\cal U}^{'}  
        \left(  
        {\hat a}^{\scriptscriptstyle \dagger}_{h}  
        {\hat a}_{k}  
        \right)  
        \psi_k  
        \rangle  
        >0  
        \qquad  
        \sum_{h_1 h_2 \atop k_1 k_2}  
        \langle  
        \psi_{h_1 h_2}  
        |  
        {\cal U}^{'}  
        \left(  
        {\hat a}^{\scriptscriptstyle \dagger}_{h_1}  
        {\hat a}^{\scriptscriptstyle \dagger}_{h_2}  
        {\hat a}_{k_2}
        {\hat a}_{k_1}
        \right)  
        \psi_{k_1 k_2}  
        \rangle  
        >0 ,  
        \end{equation}  
for any choice of  
$  
\left \{  
\psi_h  
\right \}  
$  
and  
$  
\left \{  
\psi_{h_1 h_2}  
\right \}  
$.  
\par  
The time evolution of the typical expressions (\ref{37}) can be  
studied by a procedure quite similar to that already shown in  
$\S\,2$, based on the representation (\ref{7}) in terms of the  
``superoperator'' ${\cal H}_0$,  
having  
$  
{\hat a}^{\scriptscriptstyle \dagger}_{h}  
{\hat a}_{k}  
$,  
$  
{\hat a}^{\scriptscriptstyle \dagger}_{h_1}  
{\hat a}^{\scriptscriptstyle \dagger}_{h_2}  
{\hat a}_{k_2}
{\hat a}_{k_1}
$  
as eigenstates and of the superoperator ${\cal T}(z)$, which was  
called scattering map. If suitable smoothness properties of  
${\cal T}(z)$ occur, essentially only the poles of  
$({ z - {\cal H}^{'}_0})^{-1}$  
contribute to the calculation of (\ref{7}), so that the following  
asymptotic representation holds:  
        \begin{equation}  
        \label{39}  
        {{{\cal U}^{'}(t)}}  
        \left(  
        {{{\hat a}^{\scriptscriptstyle \dagger}_{h}}{{\hat a}_{k}}}  
        \right)  
        =  
        {{{\hat a}^{\scriptscriptstyle \dagger}_{h}}{{\hat a}_{k}}}  
        +  t   {\cal L}'
        \left(  
        {{{\hat a}^{\scriptscriptstyle \dagger}_{h}}{{\hat a}_{k}}}  
        \right)  
        \qquad  
        \tau_0 \ll  t   \ll {
        \hbar  
        \over  
             |E_h - E_k|  
        }              ;  
        \end{equation}  
$\tau_0$ is linked to smoothness properties of ${\cal T}(z)$ and  
can be interpreted as the typical duration of a collision between  
two particles interacting through the potential  
$V(  
\left |  
{\mbox{\bf x}} -{\mbox{\bf y}}  
\right |  
)$;  
$\tau_0$ fixes a time scale that is assumed to be much smaller  
than the typical variation time $\tau_1$ of our relevant  
observables.  
${\cal L}^{'}$ is a linear mapping defined initially on the  
linearly independent elements  
$  
{\hat a}^{\scriptscriptstyle \dagger}_{h}  
{\hat a}_{k}  
$,  
$  
{\hat a}^{\scriptscriptstyle \dagger}_{h_1}  
{\hat a}^{\scriptscriptstyle \dagger}_{h_2}  
{\hat a}_{k_2}
{\hat a}_{k_1}
$.  
For brevity we simply describe the structure of ${\cal L}^{'}$,  
skipping the derivation.  
The formalism produces the typical structure of two-particle QM,  
with an N-body correction due to the Pauli principle. A  
two-particle scattering operator is defined by  
        \begin{equation}  
        \label{310}  
        {{\hat {\mbox{\sbold T}}}{}^{(2)}}(z)  
        =  
        {\hat V}{}^{(2)} + {\hat V}{}^{(2)}  
        {  
        1  
        \over  
        z - {{\hat H}_L}^{(2)}  
        }  
        {\hat V}{}^{(2)}_{L}  
        \qquad  
        {{\hat H}_L}^{(2)}  
        =  
        {{\hat H}^{(2)}}_0 +  
        {{\hat V}_L}^{(2)}  
          ,  
        \end{equation}  
where these operators, labelled by the index (2), are defined in  
the Hilbert space ${\cal H}^{(2)}$ of two identical particles by  
matrix elements in the two-particle (symmetric or antisymmetric)  
basis $|l_2 l_1 \rangle$; the matrix elements are:  
        \begin{eqnarray}  
        \label{311}  
        \langle l_2 l_1 | {\hat H}^{(2)}_{0} | f_2 f_1 \rangle
        &=&  
        (E_{f_1}+E_{f_2}) {1\over 2!}  
        \left(  
        \delta_{l_2 f_2}  
        \delta_{l_1 f_1}\pm  
        \delta_{l_2 f_1}  
        \delta_{l_1 f_2}  
        \right)  
        \nonumber  
        \\  
        \langle l_2 l_1 | {\hat V}^{(2)}     | f_2 f_1 \rangle
        &=&  
        V_{l_1 l_2 f_2 f_1}  
        \\  
        \langle l_2 l_1 | {\hat V}^{(2)}_{L} | f_2 f_1 \rangle
        &=&  
        (1 \pm {\hat n}_{l_1} \pm {\hat n}_{l_2}) V_{l_1 l_2 f_2 f_1}
        \end{eqnarray}  
the coefficients $E_f, V_{l_1 l_2 f_2 f_1}$ are given in
(\ref{10}) and  
(\ref{11}), the factor  
$(1 \pm {\hat n}_{l_1} \pm {\hat n}_{l_2})$  
is given in a more indirect way: the ``two-particle'' QM  
expressed by the aforementioned operators provides c-number  
coefficients in Fock-space operator expressions initially defined  
on the Fock-space basis $|\ldots { n}_f \ldots \rangle$,  
${ n}_f$ being the occupation numbers of the different field  
modes $f$: ${ n}_f \in {\rm I\!\!N}$ in the Bose case, ${  
n}_f=0,1$ in the Fermi case; therefore the factor  
$(1 \pm {\hat n}_{l_1} \pm {\hat n}_{l_2})$  
depends on the Fock-space basis elements on which the final  
Fock-space operator is acting. Also the adjoint operator in the  
two-particle Hilbert space will be useful:  
        \begin{equation}  
        \label{312}  
        {{\hat V}_R}^{(2)} =  
        {{\hat V}_L^{(2)}}{}^{\dagger},  \quad  
        {{\hat H}_R}^{(2)}=  
        {{\hat H}_L^{(2)}}{}^{\dagger},       \quad  
        \left[  
        {{{\hat {\mbox{\sbold T}}}{}^{(2)}}(z)  
        }  
        \right]^{\dagger}=  
        {\hat V}{}^{(2)} + {\hat V}{}^{(2)}_{R}  
        {  
        1  
        \over  
        z^* - {{\hat H}_R}^{(2)}  
        }  
        {\hat V}{}^{(2)}  
                      .  
        \end{equation}  
The superoperator ${\cal L}^{'}$ consists of an  
Hamiltonian  
part ${i\over\hbar}[{\hat H}_{\rm \scriptscriptstyle eff},\cdot]$ and  
of a part, analogous to the one in (\ref{tipstr}), reminding  
the Lindblad structure  
(\ref{1}).  
The formally self-adjoint Hamilton operator  
${\hat H}_{\rm \scriptscriptstyle eff}$  
is initially defined on the Fock-space basis  
$|\ldots {n}_f \ldots \rangle$  
by the following expression:  
        \begin{eqnarray}  
        \label{313}  
        {\hat H}_{\rm \scriptscriptstyle eff}  
        &=&  
        \sum_f E_f  
        {\hat a}_f^{\scriptscriptstyle\dagger} {\hat a}_f  
        + {1\over 2}  
        \sum_{l_1 l_2 \atop f_1 f_2}  
        {\hat a}_{l_1}^{\scriptscriptstyle\dagger}  
        {\hat a}_{l_2}^{\scriptscriptstyle\dagger}  
        V{}_{l_1 l_2 f_2 f_1}^{eff}
        {\hat a}_{f_2}  
        {\hat a}_{f_1}  
        \\  
        V{}_{l_1 l_2 f_2 f_1}^{eff}
        &=&  
        \langle  
               {l_2 l_1}  
        |  
        \frac 12   \!  \!  
        \left(     
        {{{\hat {\mbox{\sbold T}}}{}^{(2)}}(E_{f_1}+E_{f_2}+
	i\hbar\varepsilon)}  
        +  
        {  
        \left[  
        {{{\hat {\mbox{\sbold T}}}{}^{(2)}}(E_{l_1}+E_{l_2}+
	i\hbar\varepsilon)}  
        \right]  
        }^{\dagger}  
        \right)   
        |  
        {f_2 f_1}  
        \rangle.  
        \nonumber  
        \end{eqnarray}  
By comparison with (\ref{campo}) one can notice that introducing  
the time scale $\tau\gg \tau_0$ the coefficients  
$V_{l_1 l_2 f_2 f_1}$ related to the basic interaction between  
the field modes is replaced by  
$V{}_{l_1 l_2 f_2 f_1}^{eff}$,
linked with a full, Pauli principle corrected description of the  
two body collisions in the medium, expressed in terms of the  
self-adjoint part of the operator  
${{{\hat {\mbox{\sbold T}}}{}^{(2)}}(z)}$.  
The anti-self-adjoint part  
$\frac i2 ({{{\hat {\mbox{\sbold T}}}{}^{(2)}}(z)}  
-{{{\hat {\mbox{\sbold T}}}{}^{(2)}}(z)  
        }^{\dagger}  
)$ is not zero if one goes beyond Born approximation and provides  
a contribution to ${\cal L}^{'}$ analogous to the second term in
the l.h.s. of (\ref{1}), of the form  
$  
        - {1\over \hbar}  
        \left(  
        \left[  
        {\hat \Gamma}^{(2)} , {\hat a}^{\scriptscriptstyle\dagger}_h  
        \right]  
        {\hat a}_k  
        -  
        {\hat a}^{\scriptscriptstyle\dagger}_h  
        \left[  
        {\hat \Gamma}^{(2)}, {\hat a}_k  
        \right]  
        \right)  
$, that due to sign ``-'' cannot be rewritten as $\left[  
        {\hat \Gamma}^{(2)}, \cdot  
        \right]$;  
the operator ${\hat \Gamma}^{(2)}$ is defined on the  
Fock-space basis by  
        \[  
        {1\over 2}  
        \sum_{f_1 f_2 \atop l_1 l_2}  
        {\hat a}^{\scriptscriptstyle\dagger}_{l_1}  
        {\hat a}^{\scriptscriptstyle\dagger}_{l_2}  
        \langle  
        {l_2 l_1}  
        |  
        \frac i2   \!  \!  
        \left(     
        {{{\hat {\mbox{\sbold T}}}{}^{(2)}}(E_{f_1}+E_{f_2}+
	i\hbar\varepsilon)}  
        -  
        \left[  
        {{\hat {\mbox{\sbold T}}}{}^{(2)}}(E_{l_1}+E_{l_2}+i\hbar\varepsilon)  
        \right]^{\dagger}  
        \right)   
        |  
        {f_2 f_1}  
        \rangle  
        {\hat a}_{f_2}  
        {\hat a}_{f_1}.  
        \]  
At this point one immediately expects a third contribution to
${\cal L}^{'}$  
related to the product structure  
$  
{\hat a}^{\scriptscriptstyle \dagger}_{h}  
{\hat a}_{k}  
$  
and involving both ${{\hat {\mbox{\sbold  
        T}}}{}^{(2)}}$ and ${{{\hat {{\mbox{\sbold  
        T}}}}{}^{(2)}}}^{\dagger}$; this contribution is given by  
$  
{1\over\hbar} \sum_{\lambda}     \!
{\hat R}^{(2)}_{h \lambda}{}^{\dagger}  
{\hat R}^{(2)}_{k \lambda}  
$  
and reminds the structure of the third term at the l.h.s. of the  
Lindblad expression (\ref{1}), where it accounted for decoherence (or  
state reduction, or event production). The operators
${\hat R}^{(2)}_{k \lambda}$ are defined on the  
Fock-space basis by:  
        \[  
        {\hat R}^{(2)}_{k \lambda} =  
        -i  
        \sqrt{2\varepsilon  
        (1 \pm {\hat n}_{\lambda} \pm {\hat n}_{k} )}  
        \sum_{f_1 f_2}  
        {  
        \langle  
        {k \lambda}  
        |  
        {{{\hat {\mbox{\sbold T}}}{}^{(2)}}(E_{f_1}+E_{f_2}+
	i\hbar\varepsilon)}  
        |  
        {f_2 f_1}
        \rangle  
        \over  
        E_k +E_\lambda - E_{f_1}-E_{f_2}  
        -i\hbar\varepsilon  
        }  
        {\hat a}_{f_2}  
        {\hat a}_{f_1} .  
        \]  
The factor  
$\sqrt{2\varepsilon  
        (1 \pm {\hat n}_{\lambda} \pm {\hat n}_{k} )}$  
arises in the approximate factorisation of a Pauli correction  
term depending both on  
${\hat n}_{k}$ and  
${\hat n}_{h}$:  
        \begin{equation}  
        \label{314}  
        2\varepsilon  
        \left(  
        1 \pm {\hat n}_{\lambda} \pm \frac 12  
        ({\hat n}_{h}  
        +{\hat n}_{k})  
        \right)  
        \approx  
        \sqrt{2\varepsilon  
        (1 \pm {\hat n}_{\lambda} \pm {\hat n}_{h} )}  
        \sqrt{2\varepsilon  
        (1 \pm {\hat n}_{\lambda} \pm {\hat n}_{k} )}  
                                                 ,  
        \end{equation}  
this factorisation, which is a good approximation if the Pauli  
corrections are not very large, is a typical quantum condition,  
which together with $\tau_0 \ll \tau_1$ must be satisfied for the  
validity of the simple thermodynamic behaviour that we are  
considering in this section. The final structure of  
${\cal L}^{'}$ is formally the same as  
in (\ref{tipstr})  
        \begin{equation}  
        \label{x}  
        {\cal L}^{'}  
        {\hat a}^{\scriptscriptstyle \dagger}_{h}  
        {\hat a}_{k}  
        =  
        {i\over\hbar}  
        \left[  
        {\hat H}_{\rm \scriptscriptstyle eff},  
        {\hat a}^{\scriptscriptstyle \dagger}_{h}  
        {\hat a}_{k}  
        \right]  
        - {1\over \hbar}  
        \left(  
        \left[  
        {\hat \Gamma}^{(2)} , {\hat a}^{\scriptscriptstyle\dagger}_h  
        \right]  
        {\hat a}_k  
        -  
        {\hat a}^{\scriptscriptstyle\dagger}_h  
        \left[  
        {\hat \Gamma}^{(2)}, {\hat a}_k  
        \right]  
        \right)  
        +  
        {1\over\hbar} \sum_\lambda  
        {\hat R}^{(2)}_{h \lambda}{}^{\dagger}  
        {\hat R}^{(2)}_{k \lambda},  
        \end{equation}  
the main difference lying in the space in which these operators  
act, according to the two different physical situations.  
As a consequence of unitarity of ${\cal U}^{'}$ one can prove  
that within the approximation leading to expression (\ref{x}) one  
has:  
        \begin{equation}  
        \label{xx}  
        {\hat \Gamma}^{(2)} \approx {1\over4}  
         \sum_{h \lambda}  
        {\hat R}^{(2)}_{h \lambda}{}^{\dagger}  
        {\hat R}^{(2)}_{h \lambda} ,  
        \end{equation}  
therefore one can replace the expression  
$        {\hat \Gamma}^{(2)} \approx {1\over4}  
         \sum_{h \lambda}  
        {\hat R}^{(2)}_{h \lambda}{}^{\dagger}  
        {\hat R}^{(2)}_{h \lambda}$  
in (\ref{x}): in this way the conservation relation  
${\cal L}^{'}{\hat M}=0$ is exactly satisfied. It can be easily  
shown that  
$  
\sum_{hk}  
\langle  
\psi_h |  
(  
[  
1+\tau {\cal L}^{'}  
]  
{\hat a}^{\scriptscriptstyle \dagger}_{h}  
{\hat a}_{k}  
)  
\psi_k  
\rangle \ge 0  
$            to first order in $\tau$, so that the positivity  
property (\ref{a}) is satisfied.  
\par 
As a preliminary check of the formalism let us report that the  
calculation of  
${\cal L}^{'}  
({\hat a}^{\scriptscriptstyle \dagger}_{h}  
{\hat a}_{k})  
$            yields the typical structure of the collision term  
of the Boltzmann equation with the Pauli principle corrections:  
$  
        {1\over\hbar}  
        {\hat R}^{(2)}_{h \lambda}{}^{\dagger}  
        {\hat R}^{(2)}_{h \lambda}  
$                    and  
$  
        - {1\over \hbar}  
        \left[  
        {\hat \Gamma}^{(2)} , {\hat a}^{\scriptscriptstyle\dagger}_h  
        \right]  
        {\hat a}_h + c.c.  
$  
are respectively the ``gain'' and the ``loss'' part of the  
collision term;  
${\cal L}^{'}  
({\hat a}^{\scriptscriptstyle \dagger}_{h}  
{\hat a}_{k})  
$            yields also the streaming term  
of the Boltzmann equation.  
The study of  
${\cal L}^{'}  
(  
{\hat a}^{\scriptscriptstyle \dagger}_{h_1}  
{\hat a}^{\scriptscriptstyle \dagger}_{h_2}  
{\hat a}_{k_2}
{\hat a}_{k_1}
)  
$  
is not yet finished, it should end up with a full hydrodynamic  
and kinetic description of a one component continuum. The  
approximations leading to  
${\cal L}^{'}$  
are based on the smoothness assumption related to the condition  
$\tau_0\ll\tau_1$ and to a ``one mode'' approximation for the  
description of the dynamics in the time interval $\tau_0$. A  
finite parameter $\varepsilon \simeq \hbar\tau_0$ appears in the  
formal expression of ${\cal L}^{'}$; the final results  
do not appreciably depend on this by-product of the  
approximations if $\tau_0\ll\tau_1$.  
In a sense the simple  
thermodynamic behaviour  
expressed by (\ref{one}) arises by an approximation   and this is  
indicated by the presence of $\varepsilon$: an appreciable  
dependence of the results on $\varepsilon$ indicates a failure of  
the smoothness assumption and of the related approximations. Let  
us  stress finally  that in this approach existence of closed  
evolution equations for the thermodynamic state variables avoids  
any factorisation assumption for the distribution functions and  
therefore goes far beyond the approach based on the truncation of  
a hierarchy.  
\par  
\section{Dynamics with memory effects}  
\par         
According to $\S\,3$, when  
$ {i\over\hbar} [\hat H, \cdot] $ can be replaced by the map ${\cal L}^{'}$ 
given by (\ref{x}) the family of generalised Gibbs states ${\hat w}(t)$ 
replaces the family of statistical operators  
$\hat \varrho_t = {\cal U}({t- t_0}) \hat \varrho_{t_0}$ 
(for simplicity we assume now time-translation invariance).  
This means that to determine 
the evolution of the thermodynamic state 
from a given time $ \bar t$ onwards
 nothing else than 
$\beta$, 
$\mu $, ${{\mbox{\bf v}}}$ at that time
has to be taken into account: i.e. no bias comes by the previous 
history 
$\beta({\mbox{\bf x}},t')$, 
$\mu ({\mbox{\bf x}},t')$, ${{\mbox{\bf v}}}({\mbox{\bf x}},t')$, 
$t' < \bar t$. 
This is no longer true if the 
condition $\tau_1 \gg \tau_0$ [see eq. (\ref{39})]   is not satisfied.
Let 
us assume that at an initial time T the  statistical operator 
${\hat \varrho}_T$ can be identified with the Gibbs state ${\hat 
w}(T)$ related to it: 
        \begin{equation} 
        \label{41} 
        {\hat \varrho}_T={\hat w}(T). 
        \end{equation} 
By a straightforward calculation~\cite{Robin} one has: 
        \begin{eqnarray}
        \label{42} 
        {\hat \varrho}_t 
        &=& 
        e^{-{i\over \hbar}{\hat H}(t-
        T)} 
        {\hat \varrho}_T 
        e^{{i\over \hbar}{\hat H}(t-T)} 
        = 
        { 
        e^{
        - 
        \langle 
        \beta(T) 
        \cdot 
        {\hat e}^{(0)}[-(t-T)] 
        \rangle 
        + 
                \langle 
        [\mu(T)\beta(T)] 
        \cdot{\hat \rho}_{m}[-(t-T)]
        \rangle 
        }
        \over 
        {\mbox{\rm Tr}} \,
        e^{
        - 
        \langle 
        \beta(T) 
        \cdot
        {\hat e}^{(0)}[-(t-T)]
        \rangle
        + 
                \langle 
        [\mu(T)\beta(T)] 
        \cdot{\hat \rho}_{m}[-(t-T)]
        \rangle 
        }
        }
        \\ 
        &=& 
        { 
        e^{
        - 
        \langle 
        \beta(t) 
        \cdot{\hat e}^{(0)} 
        \rangle 
        + 
                \langle 
        [\mu(t)\beta(t)] 
        \cdot{\hat \rho}_{m}
        \rangle 
        - 
        \int_0^{t-T} d\tau \, 
        { 
        d 
        \over 
         d\tau 
        } 
        \left( 
        { 
        \langle 
        \beta(t-\tau) 
        \cdot{\hat e}^{(0)}(-\tau) 
        \rangle 
        - 
        \langle 
        [\mu(t-\tau)\beta(t-\tau)] 
        \cdot{\hat \rho}_{m}(-\tau)
        \rangle 
        } 
        \right) 
        }
        \over 
        {\mbox{\rm Tr}}   \,
        e^{
        - 
        \langle 
        \beta(t) 
        \cdot{\hat e}^{(0)} 
        \rangle 
        + 
                \langle 
        [\mu(t)\beta(t)] 
        \cdot{\hat \rho}_{m}
        \rangle 
        - 
        \int_0^{t-T} d\tau \, 
        { 
        d 
        \over 
         d\tau 
        } 
        \left( 
        { 
        \langle 
        \beta(t-\tau) 
        \cdot{\hat e}^{(0)}(-\tau) 
        \rangle 
        - 
        \langle 
        [\mu(t-\tau)\beta(t-\tau)] 
        \cdot{\hat \rho}_{m}(-\tau)
        \rangle 
        } 
        \right) 
        }
        } 
        \nonumber  
        \end{eqnarray} 
where ${\hat A}(\tau)=        e^{{i\over \hbar}{\hat H}\tau} 
           {\hat A} 
        e^{-{i\over \hbar}{\hat H}\tau} 
$ and $\langle 
\beta(t)  \cdot {\hat A} 
\rangle                = 
\int_\omega d^3\, {\mbox{\bf  x}}\, \beta(t,{\mbox{\bf x}}){\hat
A}({\mbox{\bf x}}) 
$. In the last term of the exponent the history of the 
thermodynamic state during the time interval $[T,t]$ appears; by  
${\dot  {\hat \rho}}_{m}
        = 
        -{\mbox{\rm div}}{\hat {\mbox{\bf J}}}^{(0)}_{{{m}}}$,
$ 
{\dot {\hat e}}= 
        -{\mbox{\rm div}}{\hat {\mbox{\bf J}}}^{(0)}_l 
$ it can be rewritten in the more perspicuous form: 
        \begin{eqnarray} 
        \label{43} 
        && 
        - 
        \int_0^{t-T} d\tau \, 
        { 
        d 
        \over 
         d\tau 
        } 
        \left( 
        { 
        \langle 
        \beta(t-\tau) 
        \cdot{\hat e}^{(0)}(-\tau) 
        \rangle 
        - 
        \langle 
        [\mu(t-\tau)\beta(t-\tau)] 
        \cdot{\hat \rho}_{{{m}}}(-\tau)
        \rangle 
        } 
        \right) 
        =     \hphantom{{}-{}ttttttt}
        \\ 
        && 
	\hphantom{{}-{}} 
	=
        \int_T^t d\tau' 
        \left[ 
        \langle 
        { 
        d 
        \over 
         d\tau' 
        } 
        \beta(\tau') \cdot 
        {\hat e}^{(0)}(\tau' - t) 
        \rangle 
        - 
        \langle 
        { 
        d 
        \over 
         d\tau' 
        } 
        [\beta(\tau')\mu(\tau')] \cdot 
        {\hat \rho}_{{{m}}}^{(0)}(\tau' - t)
        \rangle 
	\right. 
        \nonumber 
        \\ 
	&& 
        \hphantom{{}-{}=
        \int_T^t d\tau'[[}    
	- 
	\left. 
        \langle 
        {\mbox{\rm grad}}\beta(\tau')\cdot 
        {\hat {\mbox{\bf J}}}^{(0)}_l(\tau' - t) 
        \rangle 
        + 
        \langle 
        {\mbox{\rm grad}} 
        (\beta(\tau')\mu(\tau'))\cdot 
        {\hat {\mbox{\bf J}}}^{(0)}_{{{m}}} (\tau' - t)
        \rangle 
        \right] 
        \nonumber 
        \\ 
        && 
        \hphantom{
	{}-{}=
        }        
        + 
        \int_T^{t_0} d\tau' 
        \int_{\partial\omega} d\sigma \, {\mbox{\bf n}}\cdot 
        \left( 
        \langle 
        \beta(\tau',{\mbox{\bf x}}) 
        {\hat {\mbox{\bf J}}}^{(0)}_l({\mbox{\bf x}},\tau' - t) 
        \rangle 
        - 
        \langle 
        \beta(\tau',{\mbox{\bf x}}) \mu(\tau',{\mbox{\bf x}}) 
        {\hat {\mbox{\bf J}}}^{(0)}_{{{m}}} ({\mbox{\bf x}}, \tau' - t)
        \rangle 
        \right) 
        \nonumber  
        \end{eqnarray} 
where time and space derivatives of the thermodynamic fields 
appear on the same footing; by the last term also matter and
energy exchanges with the environment during the preparation time $[T, t_o]$
can be described. Taking expression (\ref{42}) for ${\hat 
\varrho}_t$ to calculate the basic expectation values and thus 
determining the thermodynamic state variables for $t > t_0$ as in 
(\ref{one}), one has for them closed evolution equations having 
as input 
$\beta({\mbox{\bf x}},\tau')$, 
$\mu ({\mbox{\bf x}},\tau')$, ${{\mbox{\bf v}}}({\mbox{\bf x}},\tau')$ 
$\tau' \in [T,t_0]$. Such equations are generally used and work 
under the hypothesis that  memory decays within a typical 
correlation time. This also helps to attenuate the problem of the 
initial choice (\ref{41}). We mention in passing that no problem 
about condition (\ref{41}) exists if one assumes the point of 
view of ``informational thermodynamics'': then ${\hat 
\varrho}_T={\hat w}(T)$ is just dictated by the measured values 
of the relevant variables at time $T$; however this approach does 
not explain why the previous history of a concrete collection of 
macrosystems is irrelevant just before $T$. Let us also mention 
the solution given by Zubarev: 
$T$ is shifted to $-\infty$, thus taking off any previous 
history; however this limit is highly critical and since also a 
thermodynamic limit is involved, it shifts the problem of 
thermodynamic evolution to a cosmological one. In our framework a 
time scale is associated to the system and the generator ${\cal 
L}^{'}$ should have a non-Hamiltonian part.  This provides a 
mechanism by which memory can decay. Let us assume that the 
preparation time $t_o - T$ is  larger than the decay time of 
the memory: at this point  
the history that comes before $T$ is 
irrelevant 
for the dynamics that ${\cal 
L}^{'}$ is able to describe. Then the choice (\ref{41}), that is not biased by 
this history, is adequate. 
When 
$ 
{\cal L}^{'}= {i\over\hbar} [\hat H, \cdot]  + 
{\tilde {\cal L}}^{'} 
$, calling 
$ 
{\hat \varrho}({[\beta,\mu,{\mbox{\bf v}}]},t,T) 
                                      $ 
the operator on the l.h.s. of (\ref{42})
one has: 
        \begin{eqnarray*} 
        && 
        {\mbox{\rm Tr}} 
        \left( 
        {\hat A} {\cal U}{(t,T)} {\hat \varrho}_T
        \right) 
        = 
        {\mbox{\rm Tr}} 
        \left[ 
        \left(
	T 
        e^{ 
        \int_T^t d\tau \, {\tilde {\cal L}}^{'}(\tau) 
        } 
        {\hat A} 
        \right) 
        {\hat \varrho}({[\beta,\mu,{\mbox{\bf v}}]},t,T) 
	\right]
	, 
	\\ 
        && 
        {\tilde {\cal L}}^{'}(\tau) 
        = 
        e^{{{\cal L}_0}^{'}(T-\tau)} 
        {\tilde {\cal L}}^{'} 
        e^{-{{\cal L}_0}^{'}(T-\tau)} 
	, \qquad    {{\cal L}_0}^{'}= {i\over\hbar} [\hat H, \cdot]
	.                   
        \end{eqnarray*} 
\par  
  

\vskip 20pt  
\begin{thebibliography}{99}  
  
\bibitem{GRW}  
{G.~C.~Ghirardi, P.~Pearle and A.~Rimini},  
{\it Phys.~Rev.~A}  
{\bf 42},  
{78}  
({1990}).  
\par  
  
\bibitem{Foundations}  
{G.~Ludwig},  
{\it Foundations of Quantum  
Mechanics}  
({1983}, {Springer}, {Berlin}){}.  
\par  
  
\bibitem{Davies}  
{E.~B.~Davies},  
{\it Quantum theory of open  
systems}  
({1976}, {Academic Press}, {London}){}.  
\par  
  
\bibitem{Lind}  
{G.~Lindblad},  
{\it Commun.~Math.~Phys.}  
{\bf 48},  
{119}  
({1976}).  
\par  
  
\bibitem{misure cont}  
E.~B.~Davies, {\it Commun.\ Math.\ Phys.} {\bf15}, 227 (1969); {\bf19}, 83 
(1970); {\bf22}, 51 (1971).
 A.~Barchielli, L.~Lanz, G.~M.~Prosperi, {\it Nuovo Cimento}
 {\bf 72B}, 79
 (1982); {\it Found. Phys.} {\bf13}, 779 (1983); {\it Proceedings of the
International Symposium: Foundations of Quantum Mechanics in the Light of New   
Technology} (Tokyo, 1983),  p.165;
 A.~Barchielli, G.~Lupieri, {\it J.\ Math.\ Phys.} {\bf26}, 2222 (1985);
{A.~S.~Holevo}, in  
{\it Lect.~Notes in Mathematics}  
({1988}, {Springer}, {Berlin}){, vol. 1303}, p.128;  
{\it Lect.~Notes in Mathematics}  
({1989}, {Springer}, {Berlin}){, vol. 1396}, p.229;  
L.~Lanz and O.~Melsheimer,  
in {\it Quantum Mechanics and Trajectories --  
Symposium On the Foundations of Modern Physics}, edited by   
P.~Busch, P.~J.~Lahti and P.~Mittelstaedt (World Scientific, 1993)
p.233-241; 
V.~P.~Belavkin, in A.~Blaqui\`ere (ed.), {\it Modelling and Control of 
Systems}, Lect.~Notes in Control and Information Sciences,  
({1988}, {Springer}, {Berlin}){, vol. 121}, p.245; 
V.~P.~Belavkin, {\it Phys.~Lett.} {\bf 140A}, 355 (1989);
V.~P.~Belavkin and P.~Staszewski, {\it Phys.~Lett.} {\bf
140A}, 359 (1989). 
\par  
  
\bibitem{Lanz6}  
{L.~Lanz},  
{\it Int. J. Theor. Phys.}
{\bf 33},  
{19}  
(1994).  
\par  
  
\bibitem{Lanz1}  
{L.~Lanz and O.~Melsheimer},  
{\it Nuovo Cimento}  
{\bf 108B},  
511  
(1993).  
\par  
  
\bibitem{art1}  
L.~Lanz and B.~Vacchini,  
{\it Int. J. Theor. Phys.}
{\bf 36},
67
(1997).
\par  
  
\bibitem{Erice}  
H.~Rauch, in {\it Advances in Quantum Phenomena}, edited by  
E.~G.~Beltrametti and J.-M.~L\'evy-Leblond, NATO ASI series,  
Vol.~B347, (1995, Plenum Press, New York)  
p.113{}.  
\par  
  
\bibitem{Sears}  
{V.~F.~Sears},  
{\it Neutron Optics}  
(1989, Oxford University Press, Oxford){}.  
\par   
  
\bibitem{Mlynek}  
{C.~S.~Adams, M.~Siegel, J.~Mlynek},  
{\it Phys.~Rep.}  
{\bf 240},  
{143}  
(1994).  
\par        
  
\bibitem{Vigue}  
{J.~Vigu\'e},  
{\it Phys.~Rev.~A}  
{\bf 52},  
{3973}  
(1995).  
\par         
  
\bibitem{Srinivas}  
{M.~D.~Srinivas and E.~B.~Davies},  
{\it Optica  
Acta}  
{\bf 28},  
{981}  
({1981}).  
\par  
  
\bibitem{Diosi}  
{L.~Di\'osi},  
{\it Europhys.~Lett.}  
{\bf 30},  
{63}  
({1995}).  
\par  
  
\bibitem{Lanz5}  
{L.~Lanz, O.~Melsheimer and E.~Wacker},  
{\it Physica}  
{\bf 131A},  
{520}  
({1985}).  
\par  
  
\bibitem{Holevo}  
{A. S.~Holevo},  
{\it Probabilistic and Statistical Aspects of  
Quantum Theory}  
({1982}, {North Holland}, {Amsterdam}).  
\par  
  
\bibitem{Roepke}  
{V.~G.~Morozov and G.~Roepke},  
{\it Physica}  
{\bf 221A},  
{511}  
({1995}).  
\par  
  
\bibitem{Robin}  
{W.~A.~Robin},  
{\it J.~Phys.~A}  
{\bf 23},  
{2065}  
({1990}).  
\par  
  
\bibitem{Zubarev}  
{D.~N.~Zubarev},  
{\it Non-equilibrium statistical thermodynamics,}  
(1974, Consultant Bureau, New York).  
\par  
  
\end{thebibliography}
\end{document}